\documentclass[10pt,twocolumn]{article}
\usepackage[dvips]{graphicx}
\usepackage{psfrag}
\usepackage{color}

\newcommand{\pacs}[1]{\noindent PACS: #1 \par \bigskip}

\begin{document}

\title{Chaotic Classical Scattering and dynamics in oscillating 1D potential wells.}

\author{G.A. Luna-Acosta$^1$, G. Orellana-Rivadeneyra,$^1$ A. Mendoza-Galv\'an,$^2$
             and C. Jung$^3$ \vspace{.3cm}\\
           {\it $^1$ Instituto de F\'{\i}sica, Universidad Aut\'onoma de Puebla, Apdo, Postal J-48
		 Pue. 72570, M\'exico}\\
           {\it $^2$ Centro de Investigacion y de Estudios Avanzados del IPN, Unidad Queretaro,}\\
            { \it Apdo. Postal 1-798, 76001 Quer\'etaro, M\'exico.} \\
           {\it $^3$ Centro de Ciencias F\'{\i}sicas, Universidad Nacional Aut\'onoma de
              Mexico,} \\ 
           {\it Apdo. Postal 139-B, 62191, Cuernavaca, Mexico} \vspace{.5cm}}

\date{\today
\vspace{0.8cm}
{
\vspace{0.8cm}
}\\
{
\texttt{\textbf{\LARGE Abstract}}
\vspace{0.5cm}
}\\
{
\begin{quote}
{\small We study the motion of a classical particle interacting with one, two, 
and fi\-na\-lly an infinite chain of 1D square wells with oscillating depth. For a
single well we find complicated scattering behavior even though there is
no topological chaos due to the absence of hyperbolic periodic orbits. In contrast,
for two coupled square wells there is chaotic scattering. The infinite
oscillating chain yields the generic
 transition to chaos, with diffusion in energy and in space, as the separation between wells
is increased. We briefly discuss the relevance of our results to solid state physics.}
\end{quote}
\pacs{05.45.+b, 03.20}
}
}

%
\thispagestyle{empty}

\maketitle


\section{Introduction}
 Most of our knowledge about molecular, atomic, and elementary particle physics
 comes from the analysis of scattering experiments. The potential fields
producing the scattering may be the result of one or many particles and may be autonomous
or time-dependent. Time-dependent scattering is particularly important nowadays in investigations
of transport properties  in mi\-cros\-co\-pic and mesoscopic physics, where 
sys\-tems are usually
modeled by simple Ha\-mil\-to\-nians \mbox{( see, e.g.,  \cite {Cai1}- \cite {Wenjun})}.
In this type of sys\-tems the description is generally
performed using the Schr\"odinger equation since  the devices under consideration are assumed 
to  operate predominantly in the pure quantum limit. However,  in order to get the complete picture of motion, 
it is necessary to study also their classical motion, and  the analysis of the 
quantum-classical correspondence is especially in\-te\-res\-ting when the classical motion becomes
chao\-tic under certain conditions \cite {qpaper}. 

The study of the manifestations of chaos in the transport properties of mesoscopic sytems
has yielded important results. For example, measurements of magnetoresistance in ballistic microstructures
in the shape of a chaotic billiard and circle showed clear distinctions in the power spectrum
of universal conductance fluctuations and in the line shape of the weak localization peak (\cite{Chang},\cite{Marcus}). In
\cite{luna} a criterion was proposed to distinguish between regular and chaotic dynamics by measuring
classical resistivity in a type of mesoscopic channels.
These results concern time independent phenomena.
Here we report preliminary results in the study of the classical dynamics of a type of simple
 time-dependent systems whose quantum counterparts are models of mi\-cros\-co\-pic or mesoscopic 
systems. Specifically, we consider the scattering of classical par\-ti\-cles interacting with one and two
1D squa\-re \mbox{wells} \mbox{with} periodically oscillating depth. \mbox{The} square wells  \mbox{may} re\-pre\-sent
\mbox{the} conduction band defined by a $Al_x Ga_{1-x}As-GaAs$ he\-te\-ros\-truc\-ture or a quantum dot
(see, e.g.,\cite{Leo1}, \cite{Li9})
and the oscillating depth may represent the electron-phonon interaction \cite {Phonon}
 or the presence of a monochromatic e\-lec\-tro\-mag\-ne\-tic or acoustic 
field (\cite{Berman1},\cite{Leo1},\cite{QFS}).
 We will also consider the dynamics of a classical particle interacting with a chain
of oscillating wells. Since the chain is infinite and pe\-rio\-dic the system is no longer an open system; 
the dy\-na\-mics is recurrent in phase space. This system is the classical counterpart of a particle
in a vibrating Kronig-Penney potential. The static Kronig-Penney potential in its quantum version has
been fundamental in the un\-der\-stan\-ding of solid state properties (see, e.g., \cite{Cardona}), while its classical version is trivial.
However, as will be shown below,  when the chain vibrates the classical motion becomes chaotic.

The organization of the paper is as follows. In sections II and III, respectively, we study the scattering
of particles off one and two oscillating wells. In Section IV, we consider the  oscillating chain and in 
Section V we draw conclusions.

\section{A single oscillating well}

\begin{figure}[!htb]
\begin{flushleft}
\includegraphics[scale=0.45]{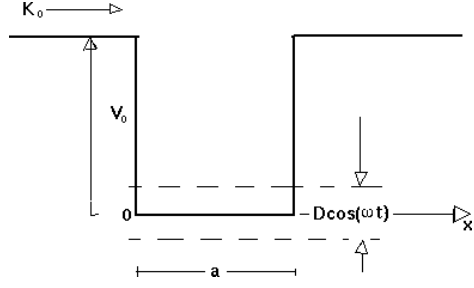}
\end{flushleft}
\caption{Schematic of the 1D oscillating well. $V_0$ is the static depth and $K_0$ is the kinetic energy of the incoming particle. }
\label{fig:1}
\end{figure}

	We consider classical particles interacting with the localized time-dependent potential, 
\begin{equation}
 V(x,t) = \left\{ \begin{array}{ll}
		V_0 & \  x < 0,  x > a \\
		D \cos (\omega t) & \  0 \le x \le a,
	    \end{array} \right.
\label{ec:1}
\end{equation}
where $a$ is the width of the well of height $V_0$, $\omega$ is the frequency of oscillation of the bottom of the well, and 
$D$ is its amplitude (See Fig. \ref{fig:1}). We shall only consider the case $D\leq V_0$.
 Assume a particle approaches the well
from the left, as indicated in Fig. \ref{fig:1}, with kinetic energy $K_0$. 
At time $t=t_0$ the particle crosses the boundary $x=0$ and the bottom of the well has a height 
$D cos(\omega t_0)$. At that instant its kinetic energy changes abruptly to $K^*= E_0 -D cos(\omega t_0)$,
where $E_0=K_0 + V_0$. 
 Since there is no gradient of $x$ within the well, the ``\mbox{}internal'' kinetic energy  $K^*$ is conserved 
while moving inside the well.
 After a time $t=t^*=a/p^*$, with $p^*= \sqrt(2K^*)$ the particle reaches the right hand side of the well ($x=a$), its total energy being
 $E_1=K^* +D cos[\omega (t_0 + t^*)]$.
If $E_1 \ge V_0$  the particle exits to the right with energy
\begin{equation}
    E_1   = E_0 + D ( cos \phi_1 - cos\phi_0), 
\end{equation}

where
	
 \begin{equation}
\phi_1  = \phi_0 + \Delta\phi_0 ,  \quad  \phi_0  = \omega t_0  \  \ and
\end{equation}

\begin{equation}
  \  \  \Delta \phi_0 \equiv  \omega t^*= \omega a/\sqrt{2(E_0- D  cos\phi_0)} 
\end{equation}

It is convenient to obtain dimensionless energies by re-scaling with respect to the height of the well $V_0$:

\begin{eqnarray}
\epsilon_1 &= & \epsilon_0 + \eta_1 (cos\phi_1 -cos\phi_0)   \\
\phi_1 & = & \phi_0 + \eta_2/ \sqrt{\epsilon_0- \eta_1 cos\phi_0} ,  
\end{eqnarray}

 where $\epsilon \equiv E/V_0$, 
\begin{equation}
 \eta_1\equiv  D /V_0, 
\end{equation}
and 
\begin{equation}
\begin{array}{rcl}
 \eta_2 & \equiv & 2\pi\omega / \omega_c,\\
 \omega_c & \equiv & 2\pi/T_c, \quad \textrm{and} \\
T_c & \equiv & a/\sqrt(2V_0).
\end{array} 
\end{equation}

Note that $T_c$ is the time for a particle with  total enery $V_0$ to cross the well in the absence of 
oscillations.

Equation (5) makes sense as long as $\epsilon_1 \geq 1$. If  $\epsilon_1<1$,  the particle reflects at $x=a$ without changing its
internal kinetic energy $K^* $. The particle then will reach $x=0$ with energy and phase given by
\begin{eqnarray}
\epsilon_2 &= & \epsilon_0 + \eta_1(cos\phi_2 -cos\phi_0) \nonumber \\
\phi_2 & = & \phi_1 + \Delta \phi_0 = \phi_0 + 2 \Delta \phi_0,            
\end{eqnarray}
where
\begin{equation}
\Delta \phi_0 = \eta_2/ \sqrt{\epsilon_0 - \eta_1 cos\phi_0}  
\end{equation}  

Again, if $\epsilon_2 < 1$ the particle will reflect towards $x=a$ and so on, until for some number $N$ of collisions with
the boundaries, the energy $\epsilon_N$ becomes greater or equal to 1. The critical value $\epsilon_N=1$ implies
that the particle exits
the well but with zero kinetic energy. Thus, the scattering map is given by
\begin{eqnarray}
\epsilon_N &= & \epsilon_0 +  \eta_1(cos\phi_N -cos\phi_0) \geq 1 \nonumber \\
\phi_N & = & \phi_0 + N \eta_2/ \sqrt{\epsilon_0 - \eta_1 cos\phi_0}            
\end{eqnarray}

This map is deceptively simple. Clearly, if $\epsilon_0 \geq  1 + 2\eta_1$ then $\epsilon_1 \geq 1$ and the particle
will certainly be transmitted 
with phase and energy given by Eq.(11) with $N=1$. How\-e\-ver, if $\epsilon_0 <1+ 2\eta_1$, then 
$\epsilon_1$ can be less 
than one, depending on the initial phase, and 
one cannot simply substitute $\phi_N$ into $ \epsilon_N$ to determine the fate of the scattered particle, i.e., 
whether it was transmitted or reflected and with what energy.
It is necessary to calculate $\phi_n$ and $\epsilon_n$ sequentially for $n=1,2,3 \ldots N$
 to dis\-co\-ver after how many collisions
$N$ the condition $\epsilon_N \geq 1$ is satisfied.

In order to get a panorama of the type of  scattering asymptotes that can be obtained for typical values of $\eta_1$
and $\eta_2$, it is useful to construct a hierarchy level plot where each level corresponds to a given number of 
collisions and the domain is the set of all in\-co\-ming asymptotes \cite{Jung1}. Such a hierarchy plot is shown in Fig. 2, for the
parameters $\eta_1=0.5$ and $\eta_2=16.45$. This plot is typical of small to moderate frequencies of oscillations
( here $\eta_2=16.45$, i.e., $\omega\approx 2.6 \omega_c$). (For higher  frequencies of oscillation of the well, the 
hie\-rar\-chy plots show increasingly complicated structures). The boundaries between the level $1$ and any other
 level $N$  are given by the set of initial conditions ($\epsilon_0, \phi_0$)  that yield outgoing energies $\epsilon_N=1$.
One may be tempted to obtain this set of ($\epsilon_0, \phi_0$) by inverting map (11) and using $\epsilon_N=1$, 
but knowledge of   the final phase $\phi_N$ is lacking. However, the following procedure enables us to obtain this set.
First we apply map (11) to the conditions $\epsilon_0=1$ and any $\phi_0$ in $-\pi\leq \phi_0\leq \pi$. 
This produces the outgoing
values $(\epsilon_n, \phi_n)$ for some integer n. Then it is easy to show, considering the symmetry of the
cosine function, that forward iteration of $(\epsilon_n, -\phi_n)$ yields the desired result 
 $\epsilon=1$. Hence, the initial conditions on the
 boun\-da\-ries, or separatrices, between hie\-rar\-chy levels are given by the set ($\epsilon_n, -\phi_n$) thus defined.
Figure 3 shows the result of this procedure for a set of 6,000  values of initial phases in the range
$-\pi\leq \phi_0\leq \pi$.\\

\begin{figure}[!htb]
\centering
\includegraphics[scale=0.45]{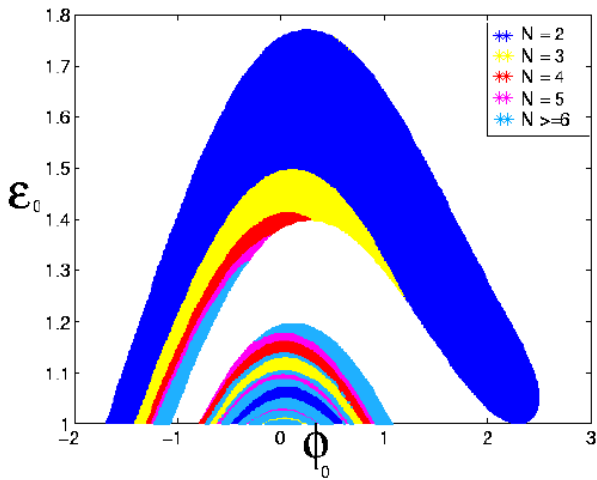}
\caption{ Hierarchy level plot \mbox{for} scat\-te\-ring \mbox{off} one oscillating square well.  \mbox{Each} co\-lo\-red a\-rea represents
the set of initial con\-di\-tions that undergo a number $N$ of collisions (transversals) within the well before exiting.
Light blue corresponds to $N$ greater or equal to 6.
The white background is the set of i\-ni\-tial conditions that transmit without any re\-flec\-tion. The parameters
are: $\eta_1=0.5$ and $\eta_2=16.45$. }
\label{fig:2}
\end{figure}

\begin{figure}[!htb]
\centering
\includegraphics[scale=0.45]{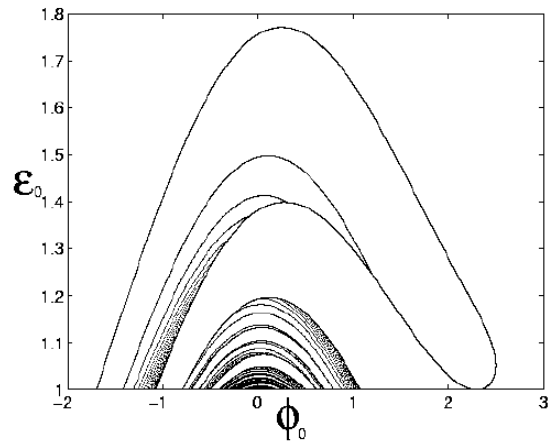}
\caption{Boundaries between hierarchy levels for the single oscillating well.  Same parameters as
in Fig. \ref{fig:2}. See text for an explanation on the construction of this plot. $\eta_1=0.5$ and $\eta_2= 16.45$.}
\label{fig:3}
\end{figure}

With the help of
 hierarchy plots like Fig. 2 one can easily determine for a given initial condition the outgoing energy and phase va\-lues, without having to iterate
the \mbox{map (11)}. One simply needs to locate the in\-co\-ming asymptotes within its hierarchy level $N$ and calculate
Eq.(11) for the known value $N$. That is, knowledge of the i\-ni\-tial con\-di\-tions e\-na\-bles us not only to determine, but
to {\it{predict}}, 
the scattering values for $N$, $\epsilon$ and $\phi$. However, for sufficiently low values of energy, the structure 
of the hierarchy levels in ge\-ne\-ral becomes  more intricate, requiring hi\-gher precision in the specification of \mbox{the} in\-co\-ming a\-symp\-to\-tes.  Figs. 4a and 4b show, res\-pec\-ti\-ve\-ly,  a plot of the number of collisions (transversals) and a plot of the final
energy as a function of $\phi_0$ for a fixed low incoming energy. The great sensitivity to initial
con\-di\-tions is further demonstrated by Figs. 5a-5d. Figure 5a is an example of how the number of
collisions can gradually increase ( here in \mbox{steps} of 2) as  $\phi_o$ is varied over smaller and 
smaller intervals and then suddenly drops to $N=1$, direct
transmission. Fig. 5b shows the co\-rres\-pon\-ding behavior of the final energy. Fig. 5c is an example of a typical situation where changes  in the initial condition
yield un\-pre\-dic\-ta\-ble, seemingly random, changes in the number of collisions and Fig. 5d is the corresponding behavior of
the final energy.

\begin{figure}[!htb]
\centering
\includegraphics[scale=0.45]{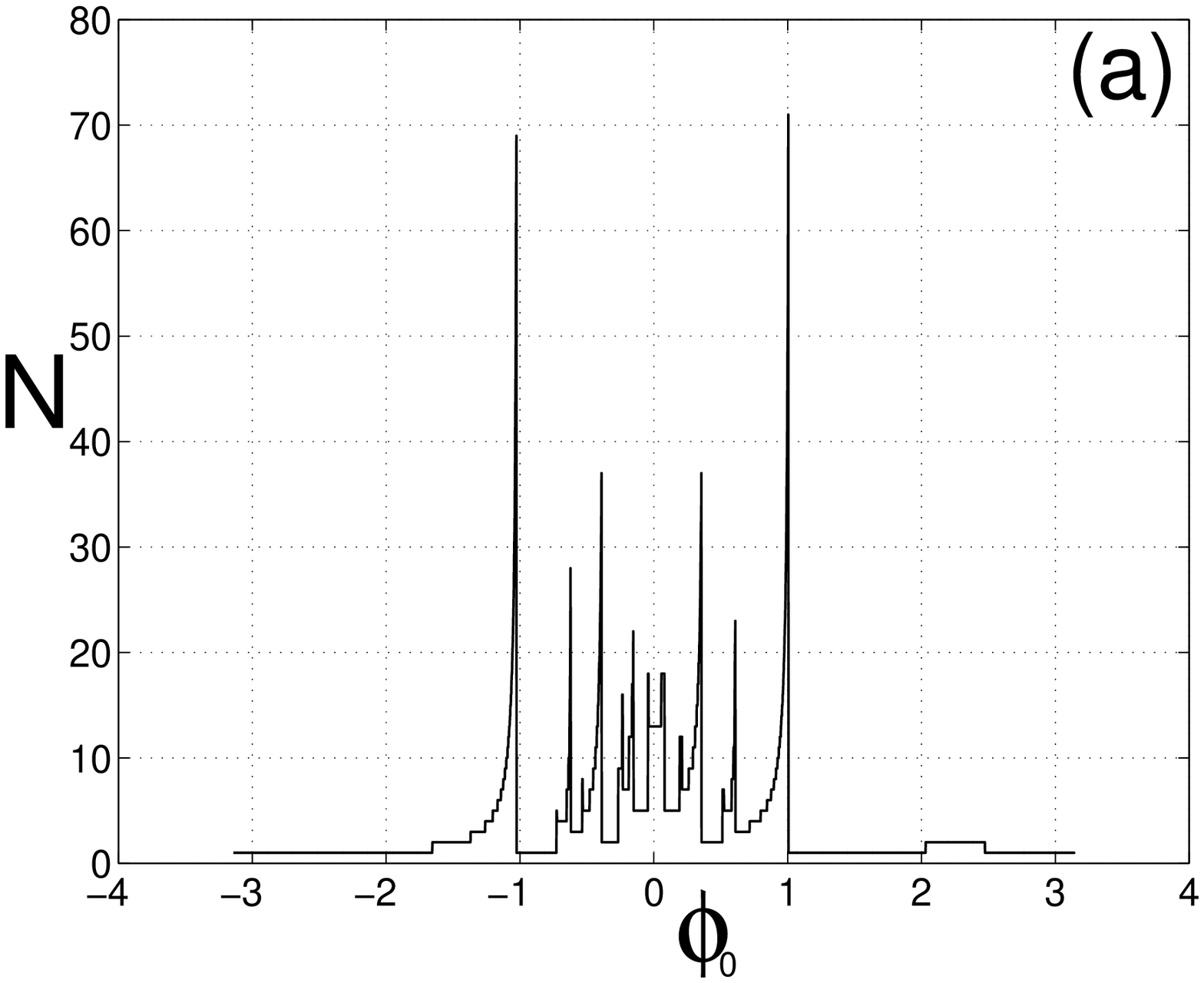}
\includegraphics[scale=0.45]{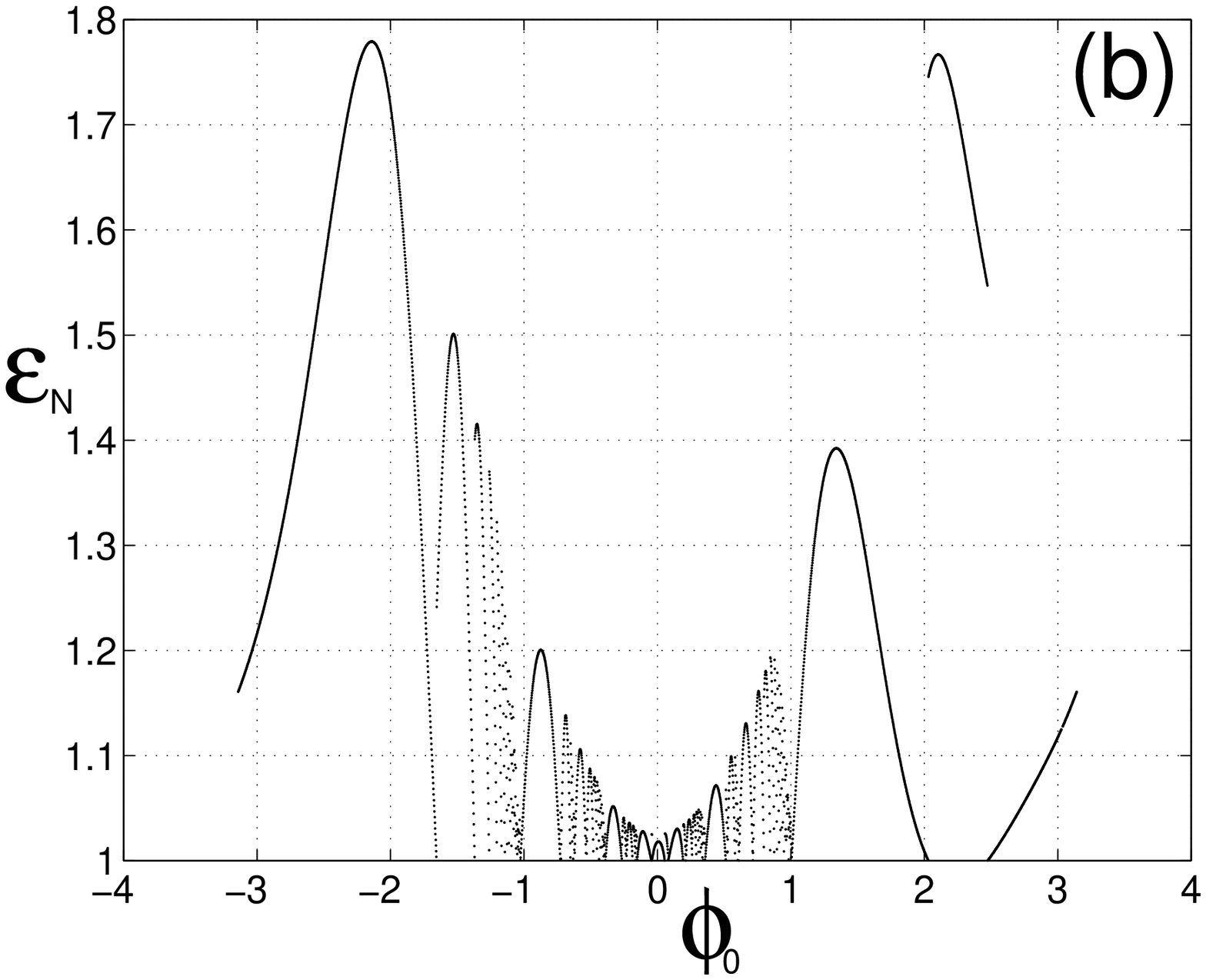}
\caption{Scattering functions for the single oscillating well ( $\eta_1=0.5,\eta_2= 16.45$). a) Number of collisions as a function
of initial phase and b) final energy as a function of initial phase. The initial energy for all phases
is $\epsilon_0 = 1.025$.}
\label{fig:4}
\end{figure}

\begin{figure}[!htp]
\centering
\includegraphics[height=5cm,width=8.2cm]{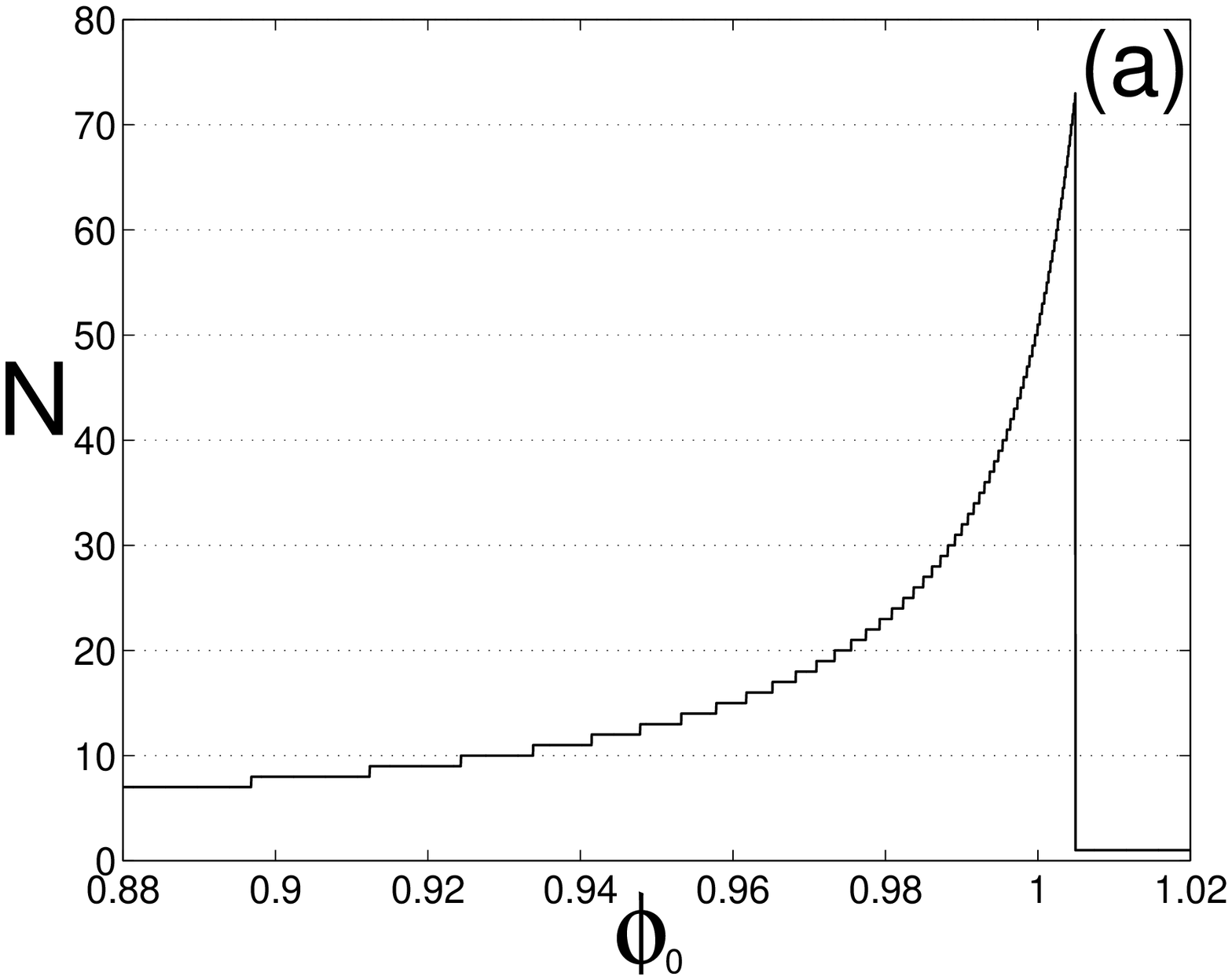}
\includegraphics[height=5cm,width=8.2cm]{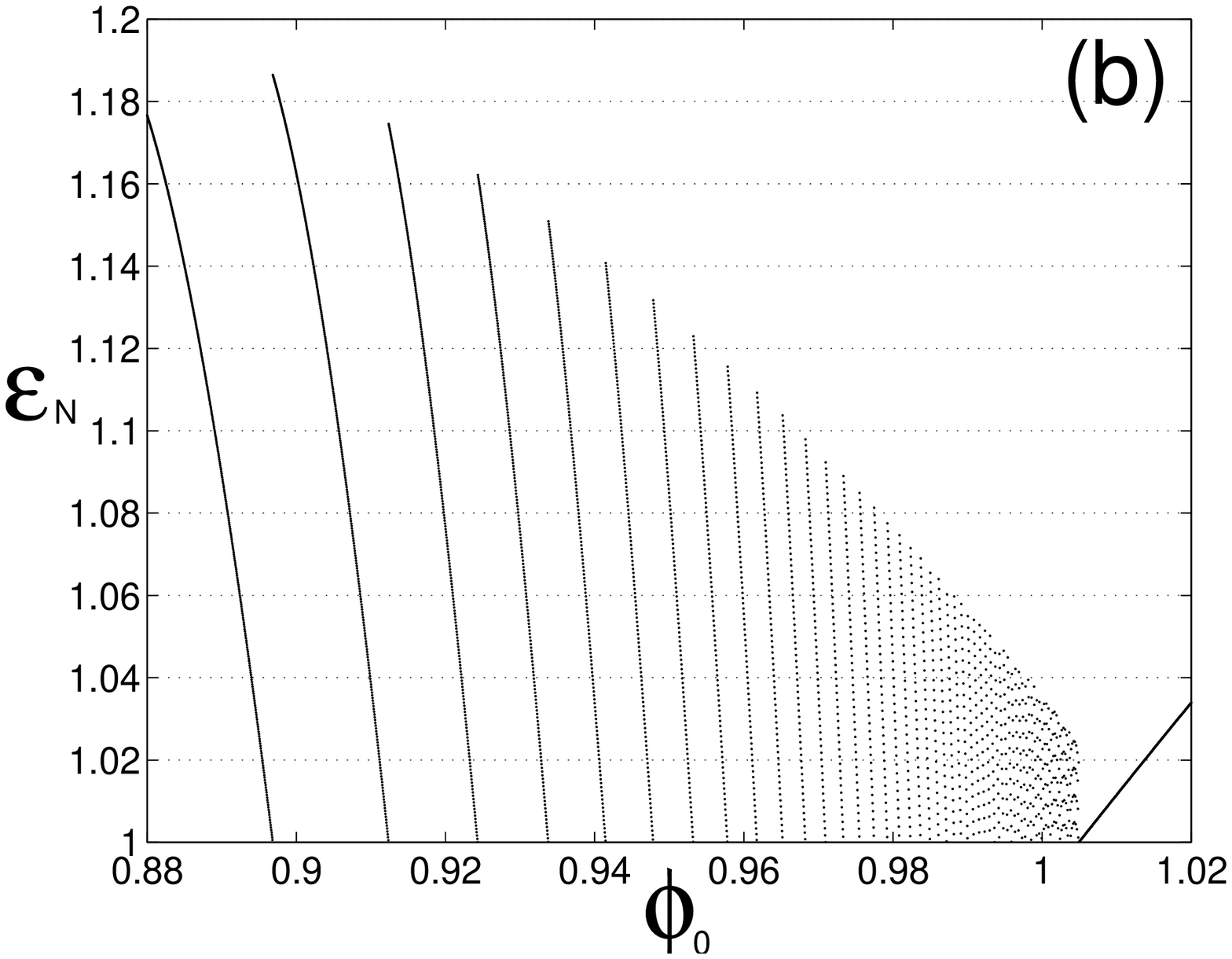}
\includegraphics[height=5cm,width=8.2cm]{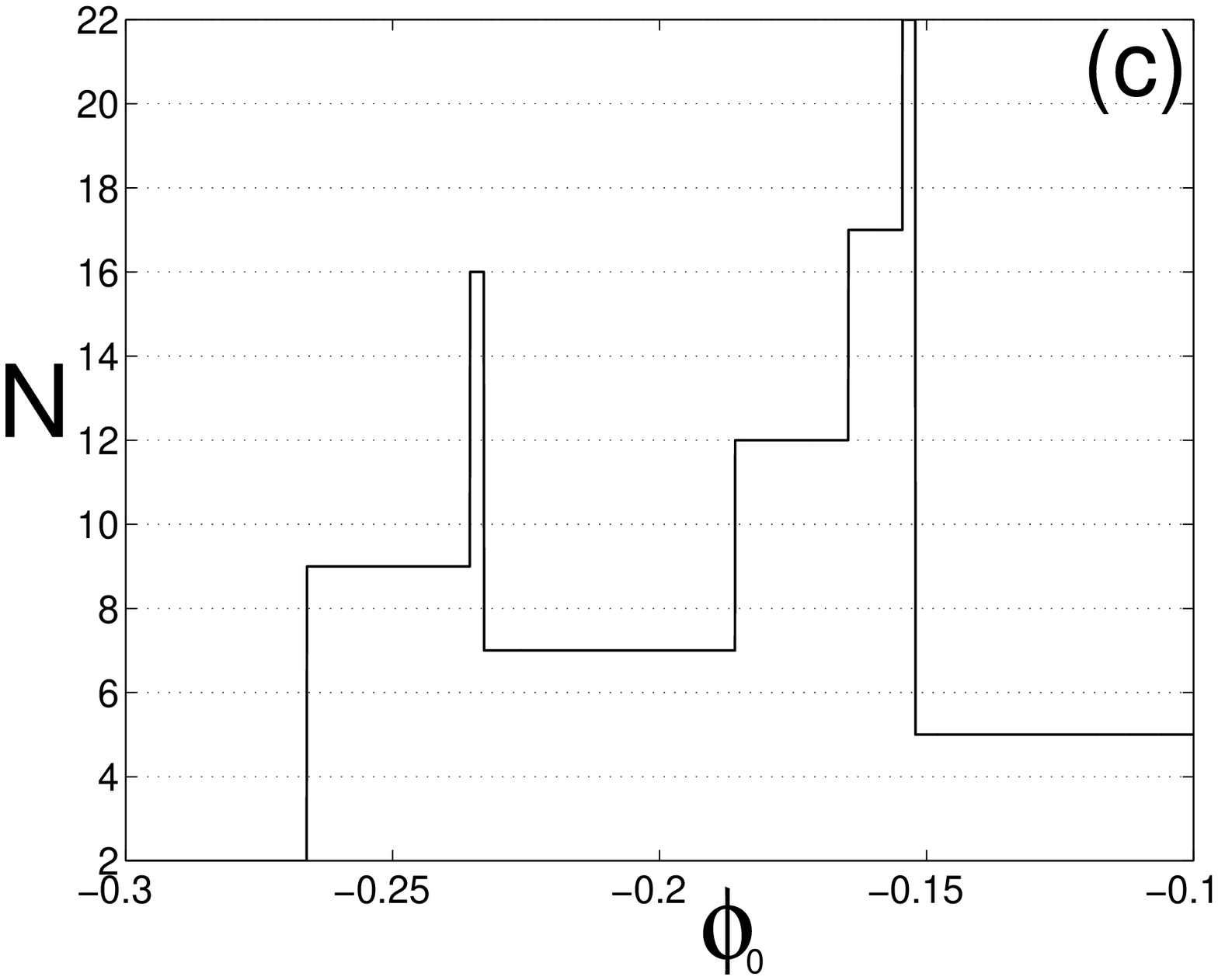}
\includegraphics[height=5cm,width=8.2cm]{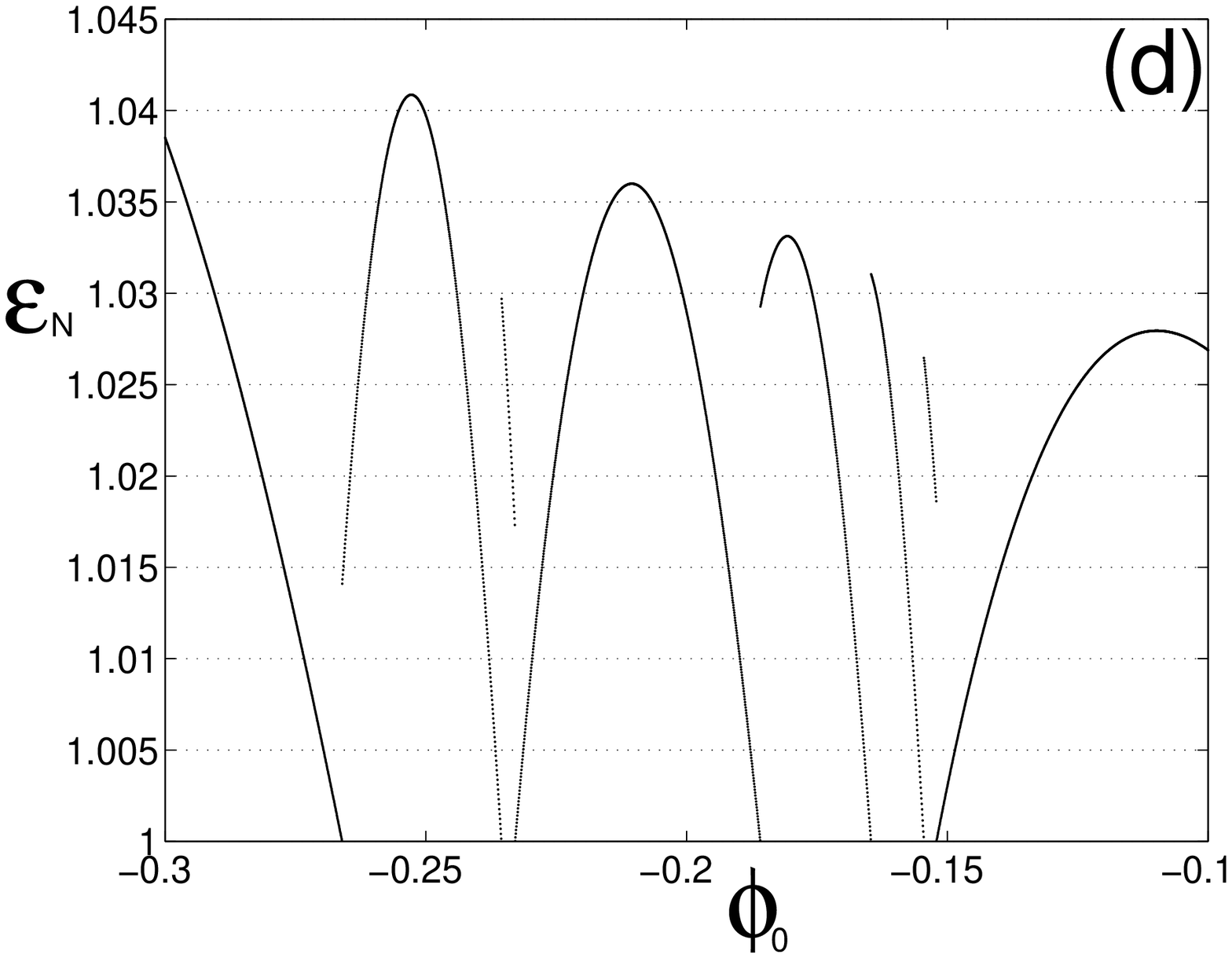}
\caption{Scattering functions for the single oscillating well ($\eta_1=0.5 , \eta_2= 16.45$) in two small ranges of initial phase. The ranges
shown here correspond to magnifications of Figures 4a and 4b. Note the two different types of behavior, namely gradual
change ((a) and (b)) and apparently random change ((c) and (d)) in the number of collisions as a function of initial phase.}
\label{fig:5}
\end{figure}

 The situation seems
even more complex when the  amplitude of oscillation is as large as the height of the 
static well, ($\eta_1=1$). Figures 6 shows the rich structure of the response functions $N(\phi_0)$ and the
outgoing energy $\epsilon(\phi_0)$, due to the extreme sensitivity to initial conditions. Notice that this  structure
is nevertheless non-fractal. Numerically, the non-fractal structure is evident because of the absence
of finer structures after increasing the resolution ( within numerical precision) enough times (see 
Fig. 7). Analytically, the fractal structure of the response ( or scattering) functions is ruled
out by realizing that there are no hyperbolic periodic points of the map.  To see this consider
the periodic orbits of period 1. The condition for period one orbits 
 ($\phi_1=\phi_0, \epsilon_1=\epsilon_0$)  gives  $ p^*= \frac{a\omega}{2\pi m}$, $m\in Z$. Within the well,
$p^{*2}/2 +\eta_1 cos(\phi)<1$, or, $cos(\phi)< \frac{1-p^{*2}/2}{\eta_1}$. Hence, one can find an
integer $m$  for a given set of parameters ($a, \omega$, and $\eta_1$) that satisfies the last inequality.
This implies that  there is  a whole interval of the angle $\phi$ that is periodic of period one.
Furthermore, since the period one points come in one-parameter families, i.e., they are not
isolated, these points cannot be hyperbolic nor elliptic. Therefore, they must be parabolic ( see, e.g.,
\cite{Tabor}). A similar argument rules out the existence of higher periodic hyperbolic points.
Topological chaos appears in systems having transverse
homoclinic or heteroclinic connections (as this implies the existence of a Smale horseshoe) which 
in turn can exist if  there are hyperbolic points \cite{Wiggins}. Consequently, there cannot be topological chaos in the case of a single well.
 Finally, if there is no topological chaos, there is no chaotic scattering.

\begin{figure}[!htb]
\centering
\includegraphics[scale=0.45]{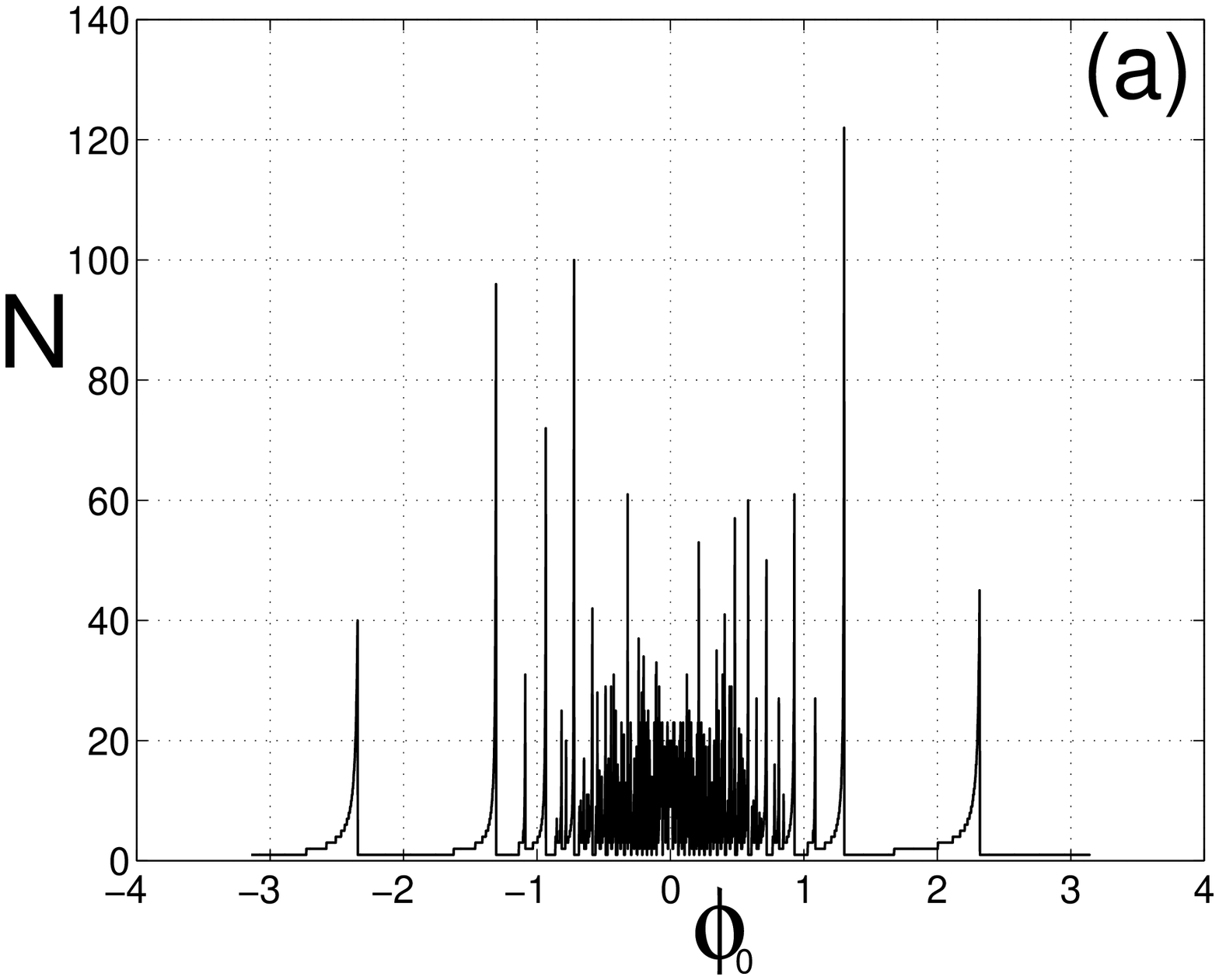}
\includegraphics[scale=0.45]{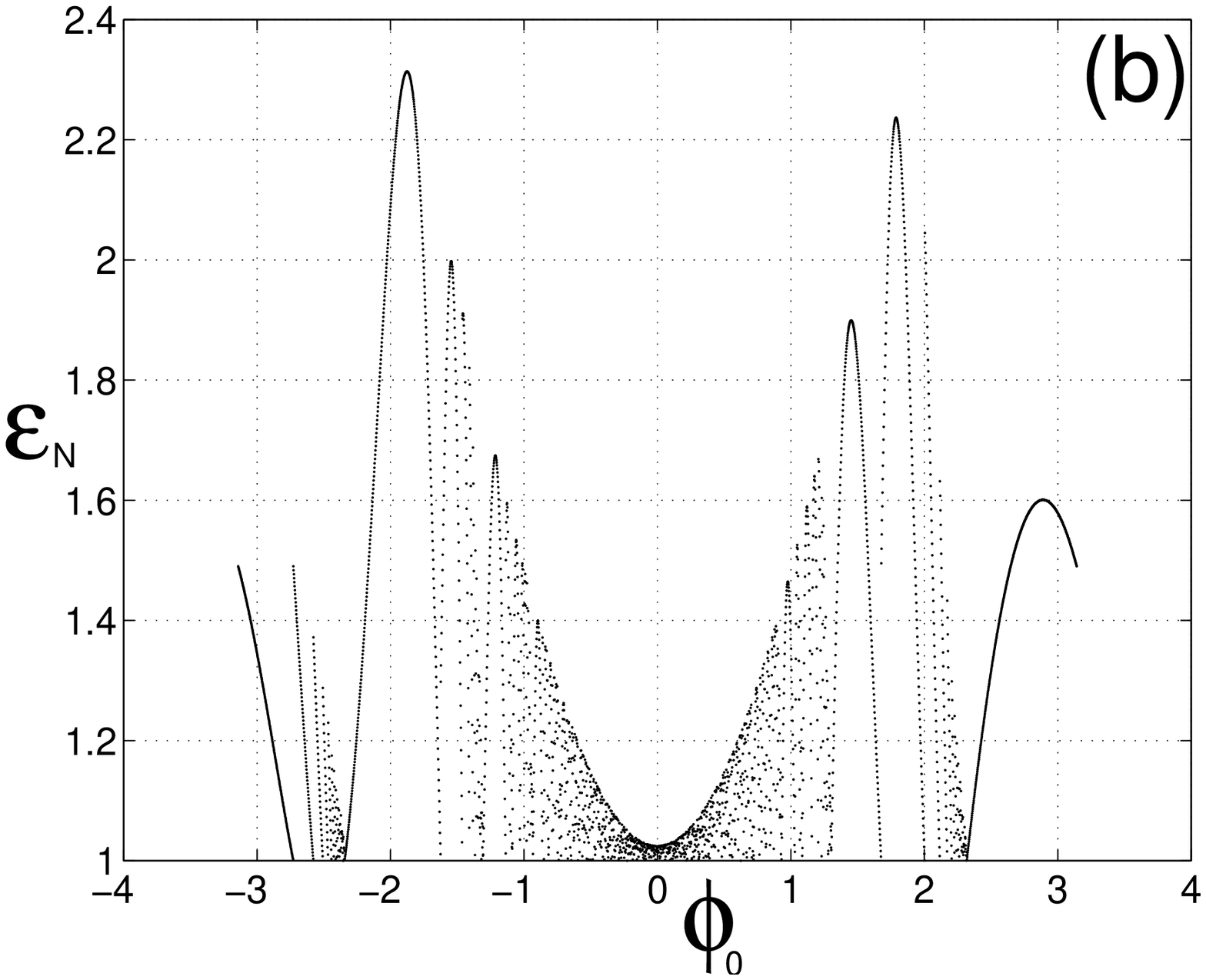}
\caption{Scattering functions for the single well with $\eta_1=1.0, \eta_2= 16.45$. a (b) is the number of collisions ( final energy) as
a function of initial phase. The initial energy for all phases is $\epsilon_0 = 1.025$. }
\label{fig:6}
\end{figure}

\begin{figure}[!htb]
\centering
\includegraphics[scale=0.45]{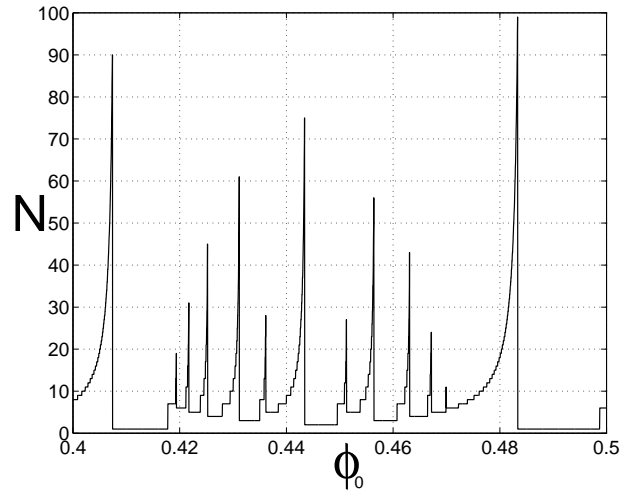}
\caption{Magnification of the scattering function of Fig. 6. }
\label{fig:7}
\end{figure}

\section{Two oscillating wells}

Now we analyze the interaction of a particle with a potential formed by two neighboring wells,
as shown in Fig. 8, oscillating in synchronization.  The separation between
the wells $b$ is a new parameter that it is convenient to make dimensionless in the same way as  was done for 
the well width $a$, namely,
\begin{equation}
\eta_3 \equiv  \omega b/\sqrt{2V_0}
\end{equation}

\begin{figure}[!htb]
\centering
\includegraphics[scale=0.45]{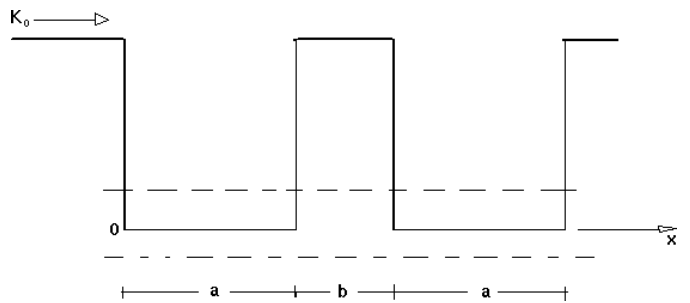}
\caption{Schematic of the oscillating double well potential.}
\label{fig:8}
\end{figure}

Assuming the particle approaches the two wells from the left and arrives at $x=0$ with energy $\epsilon_0$ and
phase $\phi_0$,  it will exit the first well with energy and phase
given by

\begin {eqnarray}
\epsilon_{m_1} &= & \epsilon_0 +  \eta_1(cos\phi'_{m_1} -cos\Phi_0) \nonumber \\ 
\phi'_{m_1} & = & \Phi_0 +   m_1 \Delta \Phi_0 ,         
\end{eqnarray}

where $m_1$ stands for the number of transversals within well number one. The reason for using the
new symbols, primed phase $\phi'_{m_1}$ and capitalized phase $\Phi_0$, will become clear shortly.
If $m_1$ is even, the particle reflects from the first well and leaves the scattering region
with energy $\epsilon_{m_1}$ but if $m_1$ is odd, the particle goes on to the second well. When it
arrives at $x=a+b$  the phase is 
\begin{equation}
  \Phi_{m_1}= \phi'_{m_1} + \delta \phi_{m_1},
\end{equation}
where
\begin{equation}
 \delta \phi_{m_1}\equiv  \eta_3 /\sqrt{\epsilon_{m_1}-1}.
\end{equation}
is the phase increase while the particle travels from $x=a$ to $x= a+b$. Note that the primed phase labels the
value of the phase right when the particle exits a well and the capitalized phase labels the phase as it enters the
well. We shall use this convention throughout the rest of the paper.
After transversing the second well $m_2$ times the particle exits with
energy and phase given by

\begin{eqnarray}
\epsilon_{m_2} &= & \epsilon_{m_1} + \eta_1(cos\phi'_{m_2} -cos\Phi_{m_1}) \nonumber \\ 
\phi'_{m_2} & = & \Phi_{m_1} +   m_2 \Delta \Phi_{m_1}.           
\end{eqnarray}

If $m_2$ is odd the particle exits the second well at $x=a+b + a$, leaving the interaction region, but
if $m_2$ is even the particle exits at $ x= a +b$, returning toward the first well, and so on.  Let $m_i$ stand for the number of
transversals within the well number $i$, where $i$ odd (even) refers to well number one ( two).
 We can then write

\begin{eqnarray}
\epsilon_{m_n} &= & \epsilon_{m_{n-1}} +\eta_1(cos\phi'_{m_n} -cos \Phi_{m_{n-1}}) \nonumber \\ 
\phi'_{m_n} & = & \Phi_{m_{n-1}}+   m_n \Delta \Phi_{m_{n-1}},
\end{eqnarray}

where
\begin{equation}
\Delta \Phi_{m_i}  \equiv \eta_2/\sqrt{\epsilon_{m_i}-\eta_1 cos \Phi_{m_i}}    
\end{equation}
and $m_i$ is even for all $i$ except $i=1$ and $i=n$ ( The case $m_1$= even is given by 
Eq. (13), where the particle did not interact with the second well). $n$ labels the last well the particle
crossed before leaving the interacting region  and so is also the total number  of interactions with the
wells. On the other hand, the total number $N$ of transversals is the sum of all $m_i$ from $i=1$ to
$i=n$.   
Figure 9 is the hierarchy plot  computed for $\eta_3=5$ and with $\eta_1$ and $\eta_2$ the same
as for figure 2. Note that for two wells,  a total number of $N=2$ transversals  can be produced in two
 ways: either, two transversals in 
the first well with no interaction with the second well, or one transversal in  the first well and one
with the second well. Figure 9 associates the color blue to the latter possibility while the white area
corresponds to the first possibility. Comparison with Fig. 2 shows that the smooth boundaries between
various hierarchy levels now appear fragmented and the scattering functions are expected to show
some evidence of this fragmentation.  Figures 10a and 10b plot the energy and the number of collisions,
respectively, as a function of the initial phase for the same parameters as those of Fig.9 and for an
incoming energy $\epsilon_0 =1.025$.  At first sight the plots $N(\phi_0)$ do not appear qualitatively different from
those corresponding to a single well (Figs. 4 and 6) but a closer look reveals that as the resolution is increased there
appears more and more structure, suggesting its fractality ( see Fig. 11).  That such fractality exists 
can be demonstrated by considering the dynamics of particles initially placed between the wells.
Let us first define the Poincare surface to be the midpoint between the two wells and positive momentum and choose
the  coordinates to be the energy and phase. The energy at the Poincare surface is given by the first of
Eqs. (17) but the phase is shifted by half of $\delta \phi_{m_n}$ to correspond to the midpoint between the wells. Thus, the
Poincare map is
\begin{eqnarray}
\epsilon_{m_n} &= & \epsilon_{m_{n-1}} +\eta_1(cos\phi'_{m_n} -cos \Phi_{m_{n-1}}) \nonumber \\ 
\phi_{m_n} & = & \phi'_{m_n} + \frac{1}{2} \delta \phi_{m_n}.    
\end{eqnarray}

\begin{figure}[!htb]
\centering
\includegraphics[scale=0.45]{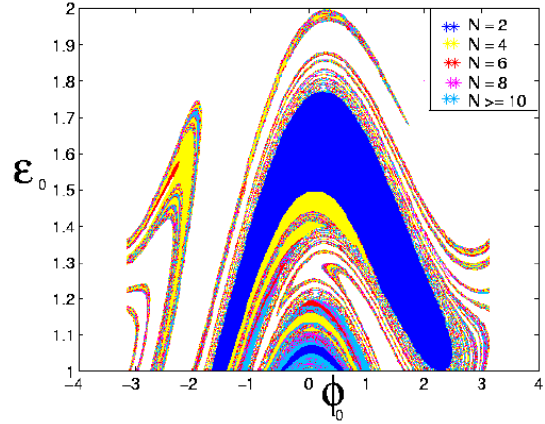}
\caption{Hierarchy level plot for the oscillating double well. There are two ways of producing $N=2$ as mentioned
in the text. The white area corresponds to direct transmission (one transversal in the first well and one in
the second one). The blue area corresponds to two transversals in the first well and none in the second one. 
 In general, there are $2^{N/2}$ ways. Here all possibilities are
plotted for $N$ equal or greater than 4. The parameters are $\eta_1=0.5$, $\eta_2=16.45$, and $\eta_3=5$. }
\label{fig:9}
\end{figure}

\begin{figure}[!htb]
\centering
\includegraphics[scale=0.45]{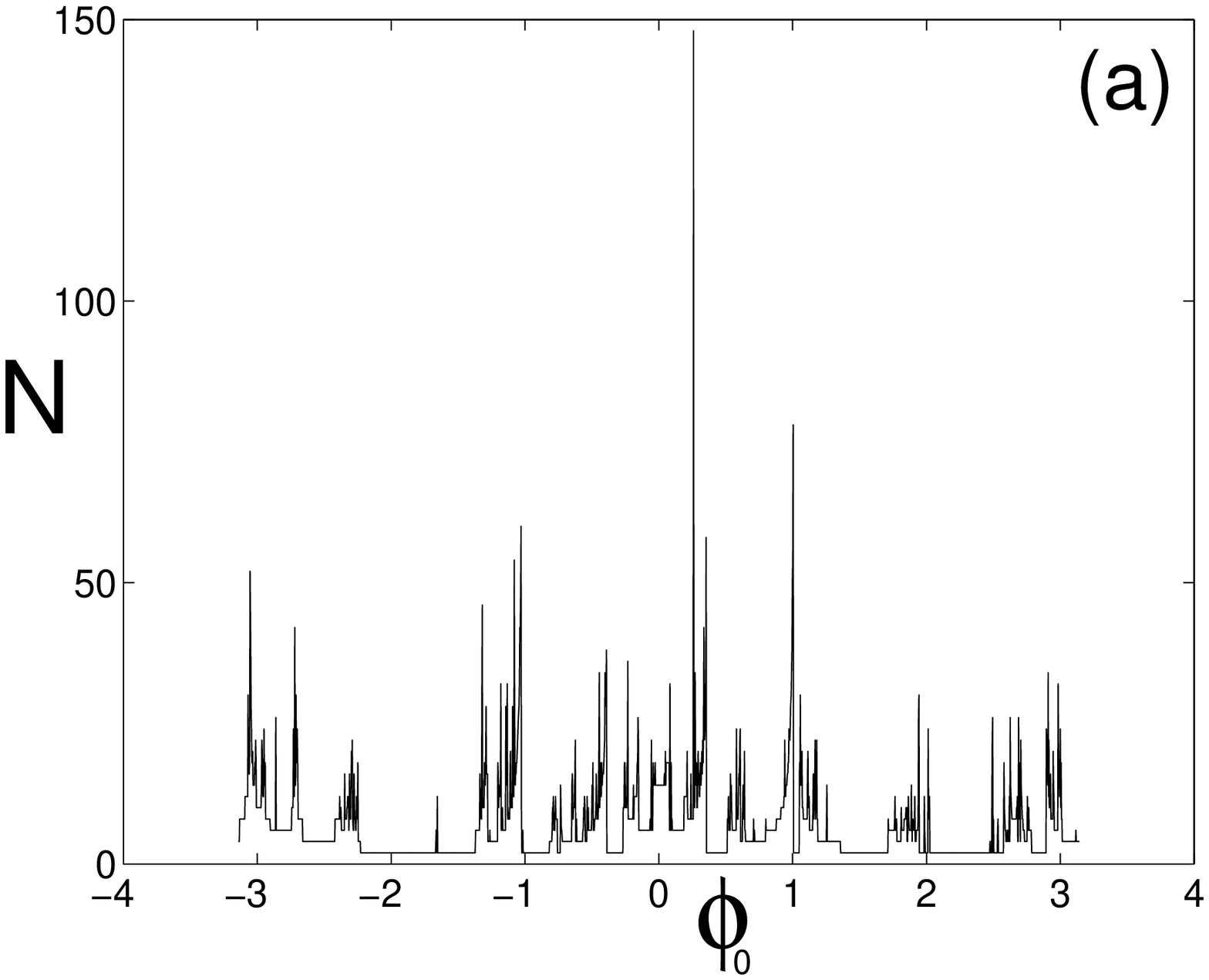}
\includegraphics[scale=0.45]{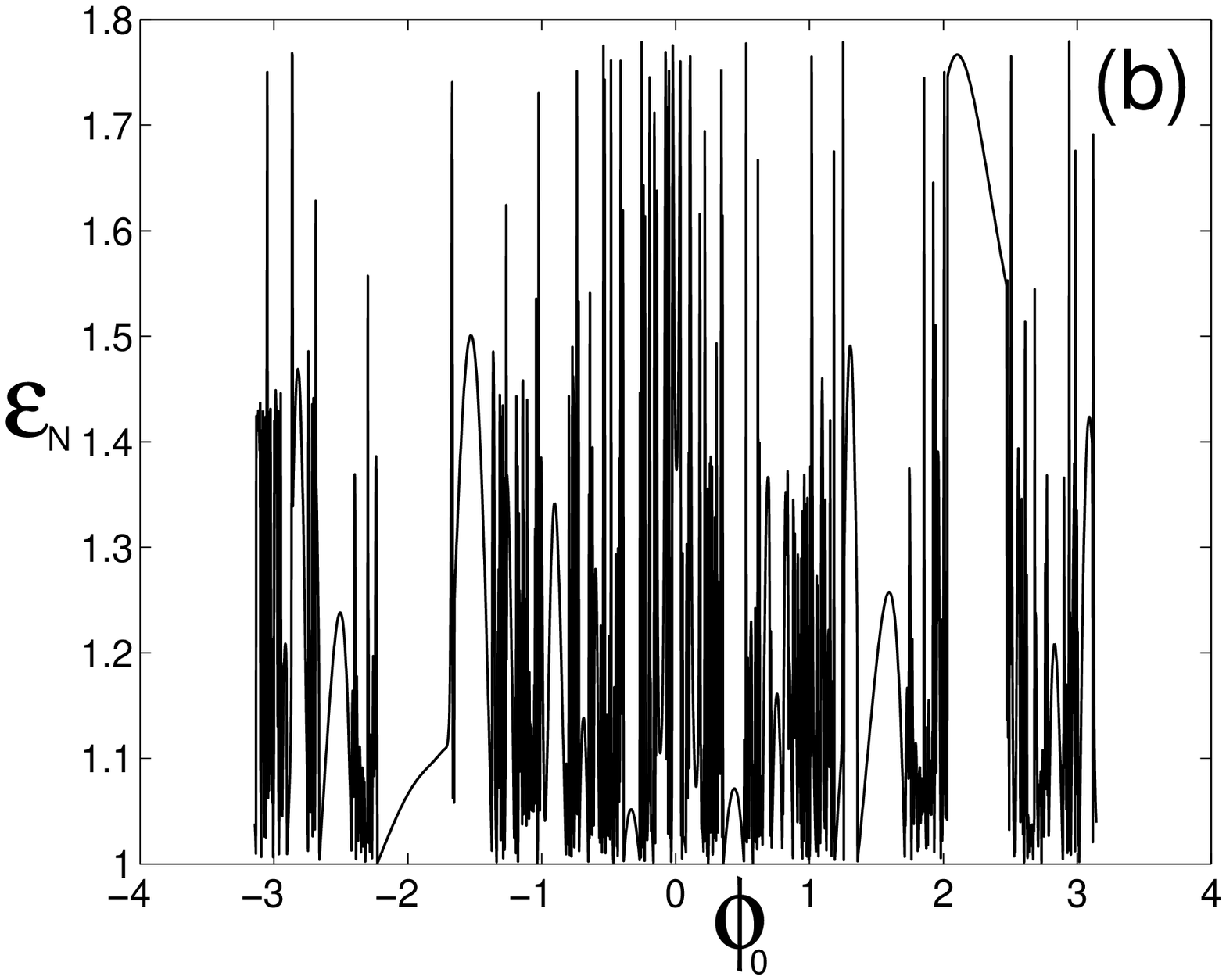}
\caption{Scattering functions for the oscillating double well. a) Number of collisions as a function
of initial phase and b) final energy as a function of initial phase. The initial energy for all phases
is $\epsilon_0 = 1.025$. $\eta_1=0.5$, $\eta_2= 16.45$, and $\eta_3=5.0$ }
\label{fig:10}
\end{figure}

\begin{figure}[!htb]
\centering
\includegraphics[scale=0.45]{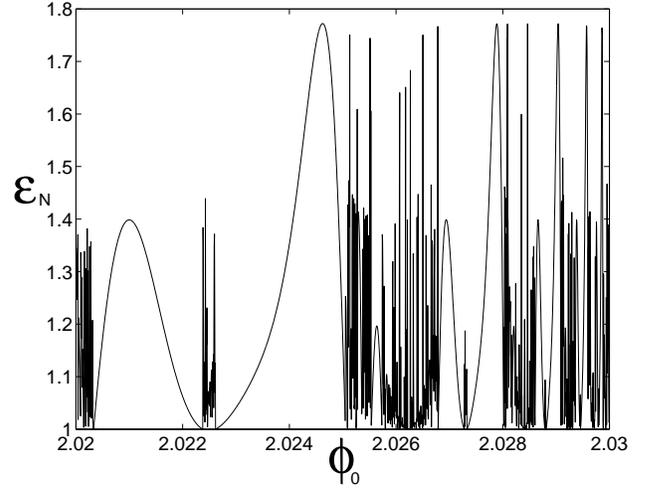}
\caption{Magnification of Figure 10b. }
\label{fig:11}
\end{figure}

 We now place a set of initial conditions on a grid on this surface and let them evolve forward in time.
 Some initial
conditions will leak after a finite number of interactions with one or  both wells and some will remain trapped
forever.  Fig.12a is a plot  for those orbits that return to  the Poincare surface at least once and up to 1000 times. 
In Fig. 12b we plot only those orbits that returned to the surface at least 100 times and up to 1000 times. That is,
transient orbits that crossed the Poincare surface less than 100 times were not plotted in Fig. 12b.
Finally, in  Fig. 13 we plot orbits that cross this Poincare section
at least 4000 times and up to 5000 times. This Poincare plot now represents the dynamics
of the particles that are trapped forever (bounded orbits)  and  shows the appearance of an island of stability associated with a
fixed point of period one and the surrounding elliptic islands of higher period ( here period 8). This is just the common structure
predicted by the Poincare-Birkhoff theorem \cite{LL} with alternating elliptic and hyperbolic fixed points. The hyperbolic
points 
and their associated homoclinic/heteroclinic tangle are not shown in this figure. The outermost island
chains are not screened behind a KAM curve, therefore the scattering flow can reach the homoclinic
tangle. Correspondingly, the tendrils of the stable and unstable manifolds of this tangle reach out 
into the asymptotic region, thus producing the fractal structure of the scattering functions \cite {Chaos3}.
 Hence, we have a numerically based proof that 
a particle scattering of\mbox{}f this double well oscillating potential gives rise to true chaotic scattering, unlike the case of a
single well.

\begin{figure}[!htb]
\centering
\includegraphics[scale=0.45]{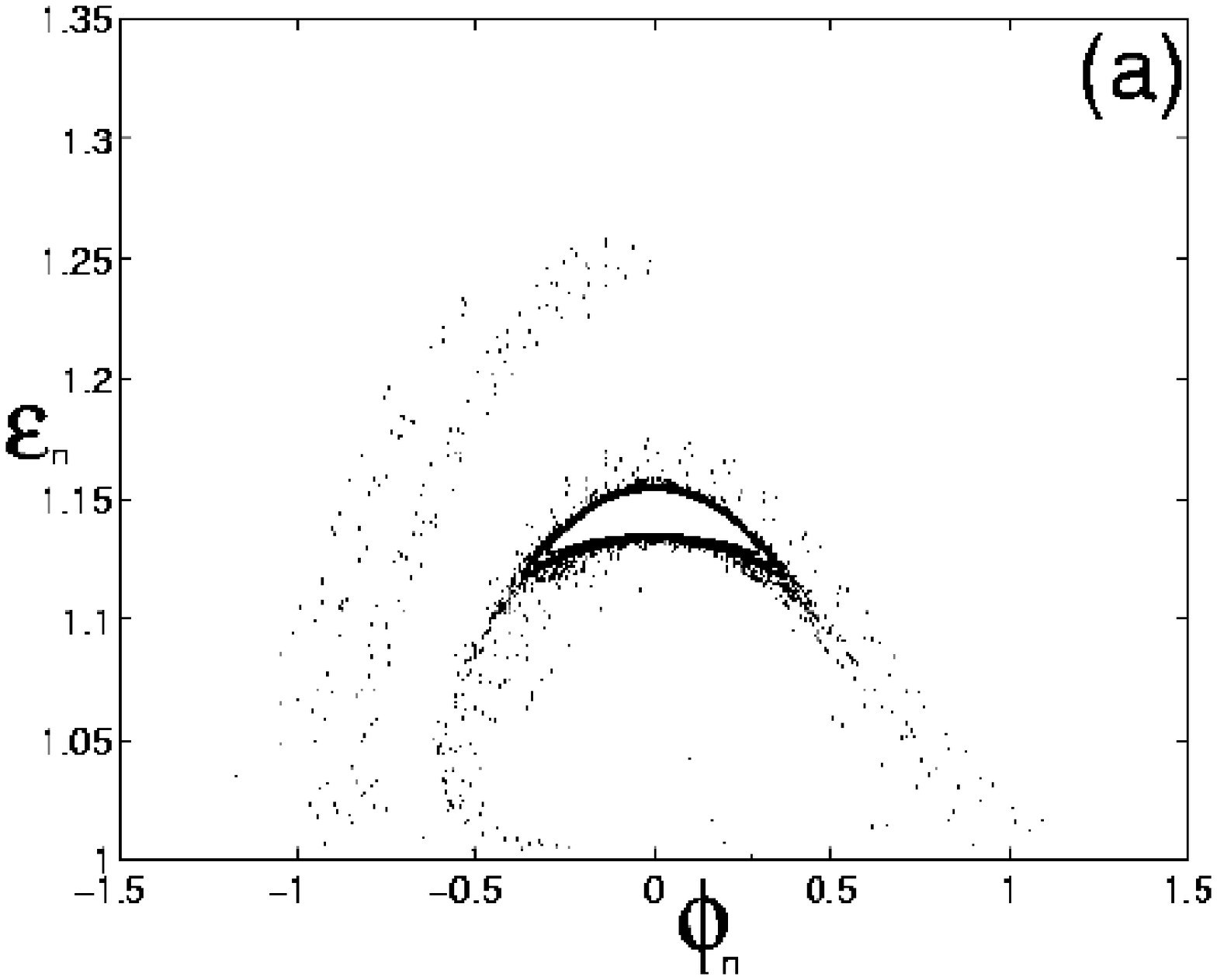}
\includegraphics[scale=0.45]{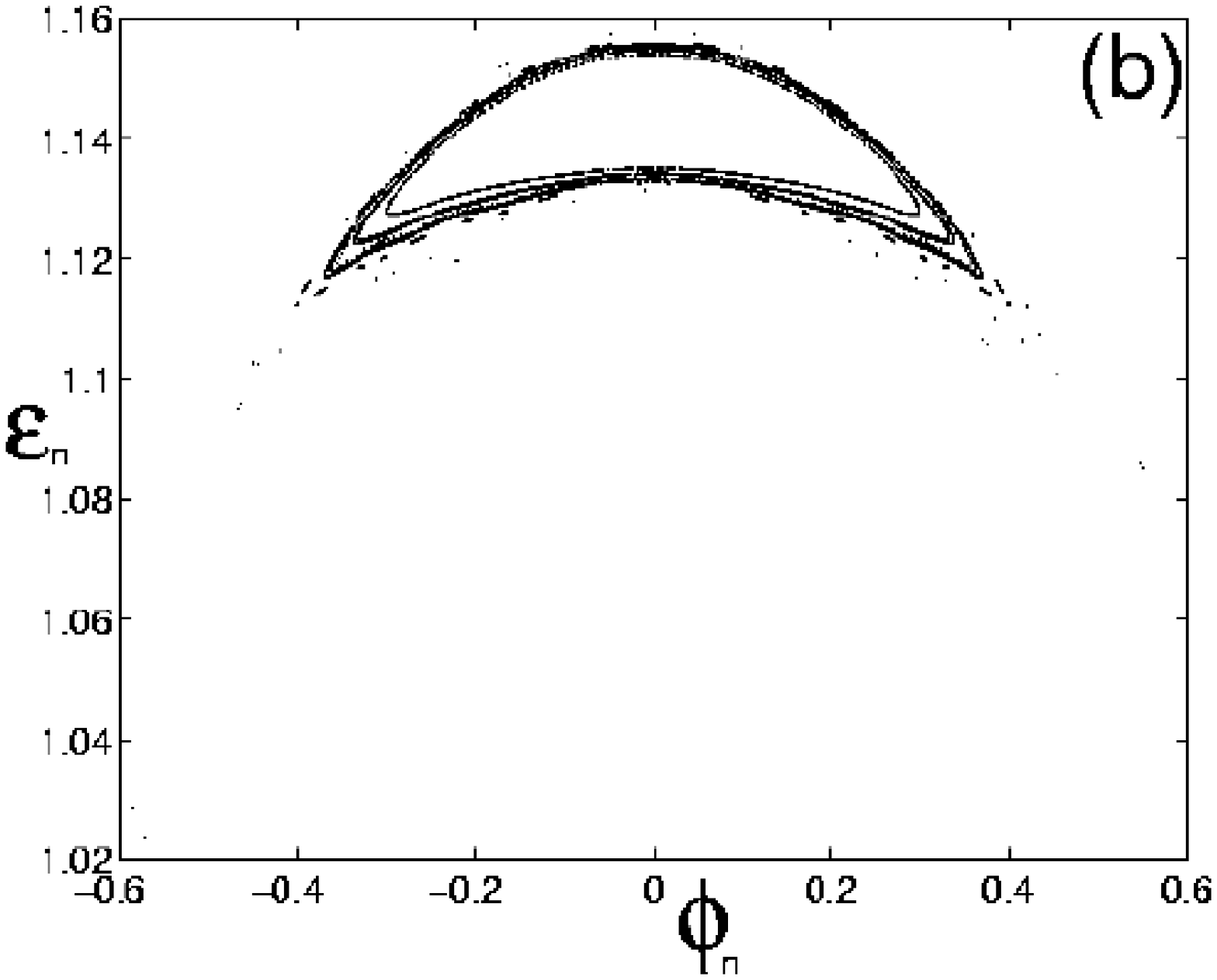}
\caption{Transient and bound orbits for the oscillating double well. 
 a) is the Poincare plot  for those
orbits that return to the surface once and up to 1000 times; b) same as a) but plots only those that crossed the
 surface at least 100 times and up to 1000 times.  Here $\eta_2= 6.28$, $\eta_3= 0.1$, and $\eta_1=0.5$ }
\label{fig:12}
\end{figure}

\begin{figure}[!htb]
\centering
\includegraphics[scale=0.45]{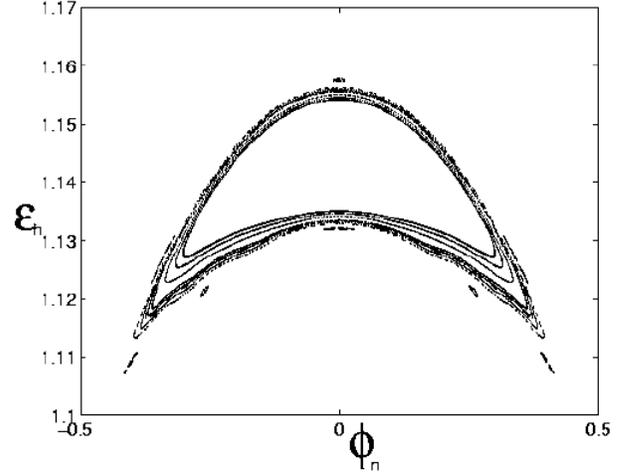}
\caption{Poincare plot for the oscillating double well. 
Here the minimum number of crossings can be as large as 5000 for some orbits. Orbits that left the interactions
region before 4000 crossings are not plotted. Same parameters as Fig.12. }
\label{fig:13}
\end{figure}

\section{Oscillating chain}

The infinite oscillating chain is formed by the infinite repetition of the double well arrangement and hence
(19) is the Poincare map describing the motion of the particle,  except that $m_i$ has
a different meaning now. Namely, $m_i$ now stands for $m_i$ transversals during the $ith$ interaction of the particle
with a well. Of course, here  $m_i$ can be odd or even for {\bf any} $ i$, $i=1,2,3....$.  Since the
chain is periodic the Poincare plots are modulo $a+b$. Figures 14-17 show the Poincare
plots for increasing values of $ \eta_3$ ($\eta_1=0.5; \eta_2=16.45$). In these figures the phase points  have been labeled $(\epsilon_n, \phi_n)$
, $n=1,2,3...$ instead of $(\epsilon_{m_n},\phi_{m_n})$, $n=1,2,3...$, because these plots
do not contain the information on how many transversals $m_n$ occur for each interaction $n$.

For the analysis of these plots it is convenient to write the energy $\epsilon_{m_n}$ of eq. (19) in terms of the
initial conditions:
\begin{eqnarray}
  \epsilon_{m_n} & = & \epsilon_0 + \eta_1( cos\phi'_{m_n} -cos\Phi_0) + {}
			\nonumber\\
& & {} + \eta_1 \sum_{i=1}^{n-1}( cos\phi'_{m_i} -cos\Phi_{m_i}) \nonumber\\
  \phi_{m_n} & = & \phi'_{m_n} + \frac{1}{2} \delta \phi_{m_n}.
\end{eqnarray}

\begin{figure}[!htb]
\centering
\includegraphics[scale=0.45]{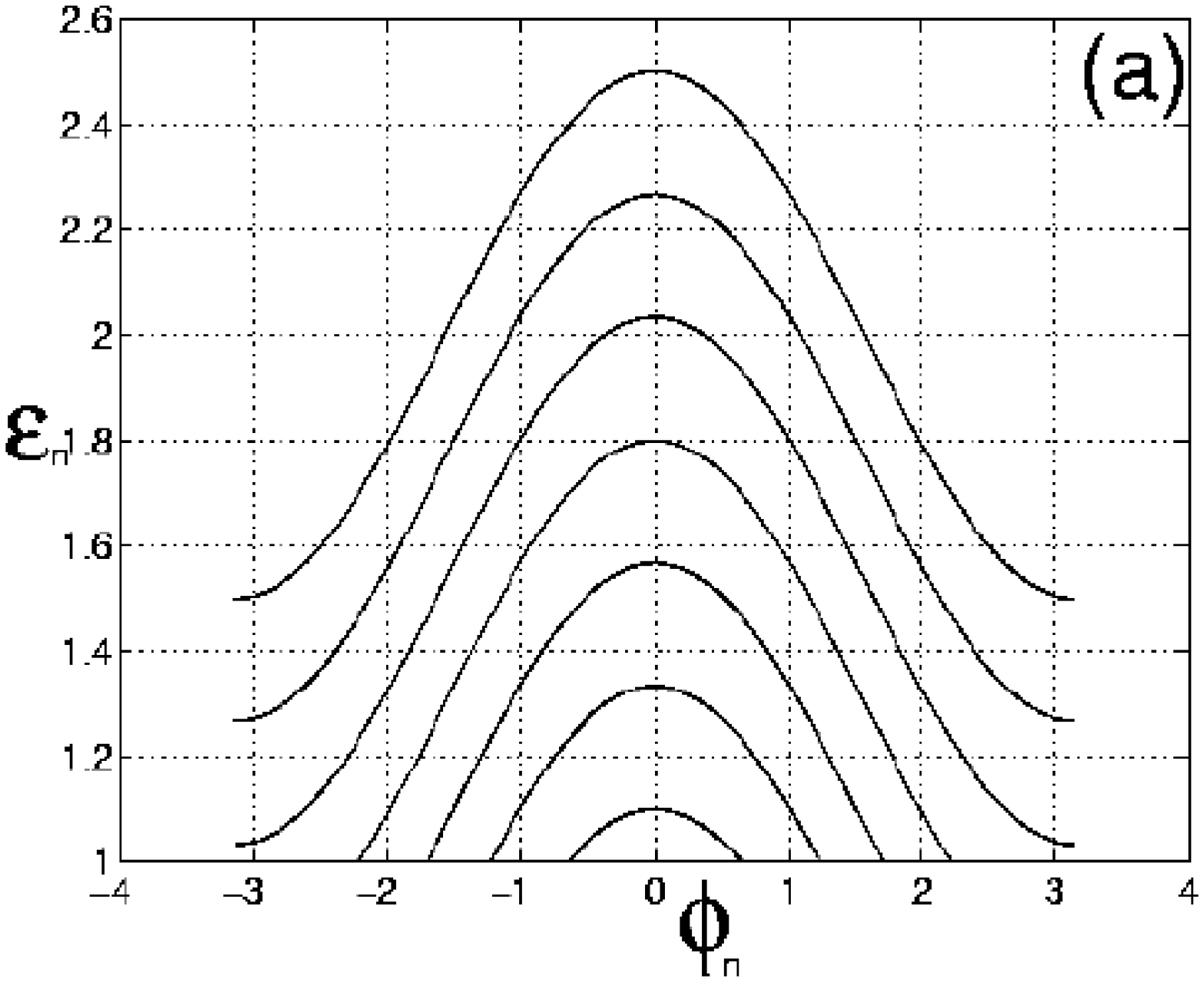}
\includegraphics[scale=0.45]{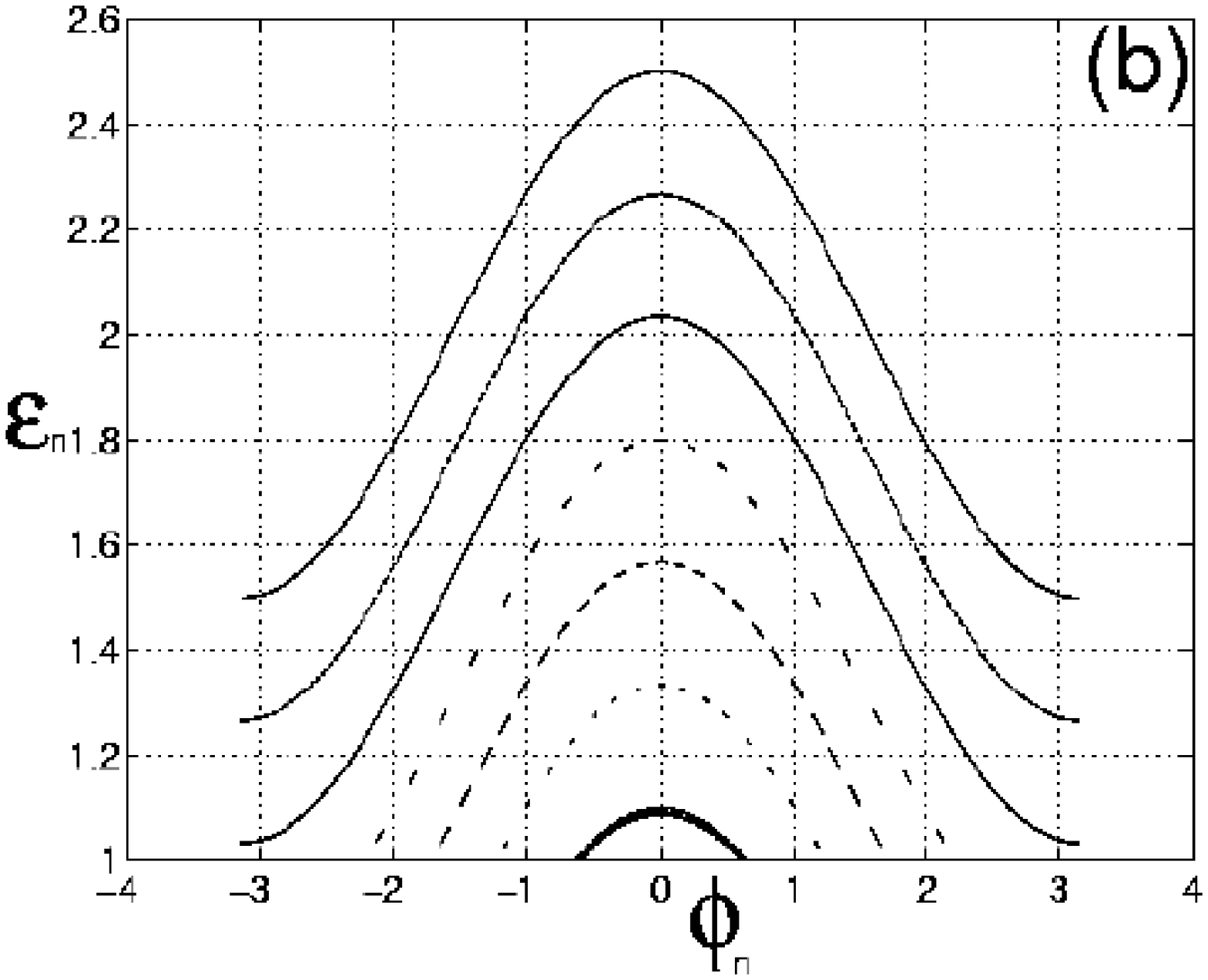}
\caption{Poincare plots for the oscillating chain of wells. a) $\eta_3=0$ and b) $\eta_3=0.0001$. $\eta_1=0.5$ and
$\eta_2= 16.45$.}
\label{fig:14}
\end{figure}

\begin{figure}[!htb]
\centering
\includegraphics[scale=0.45]{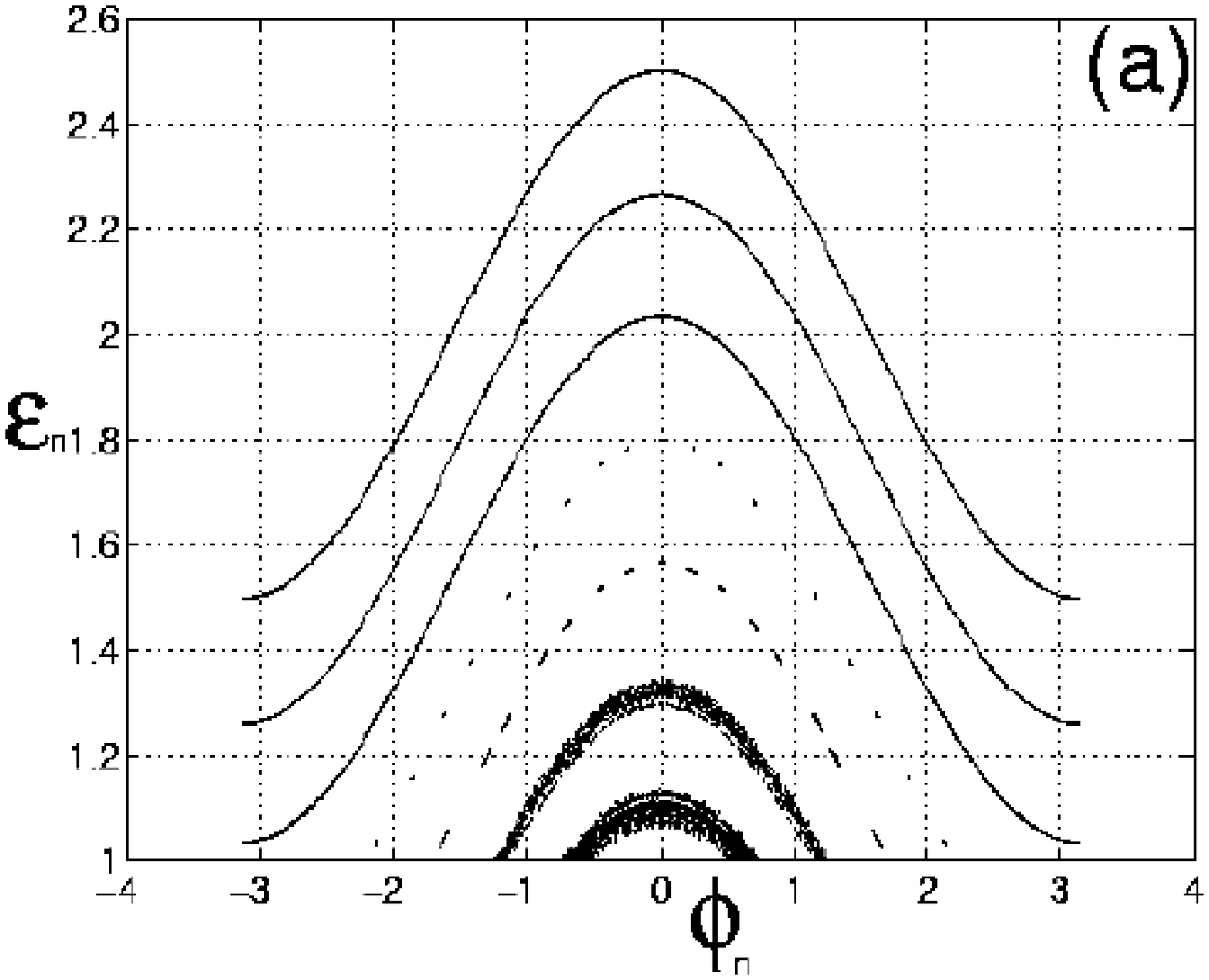}
\includegraphics[scale=0.45]{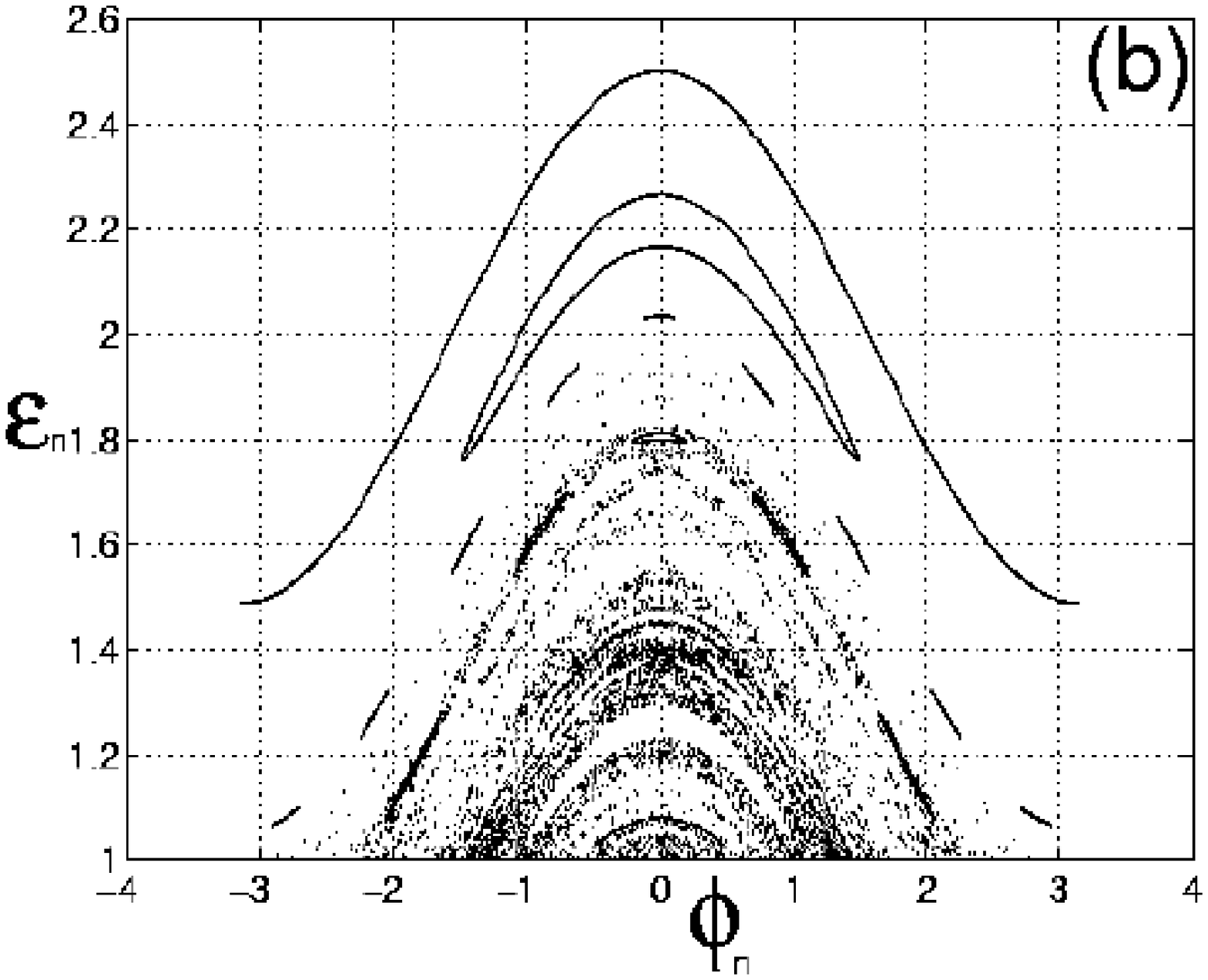}
\caption{Poincare plots for the oscillating chain of wells. a) $\eta_3=0.001$ and b) $\eta_3=0.01$. Same $\eta_1$ and
$\eta_2$ as in Fig. 14. }
\label{fig:15}
\end{figure}

\begin{figure}[!htb]
\centering
\includegraphics[scale=0.45]{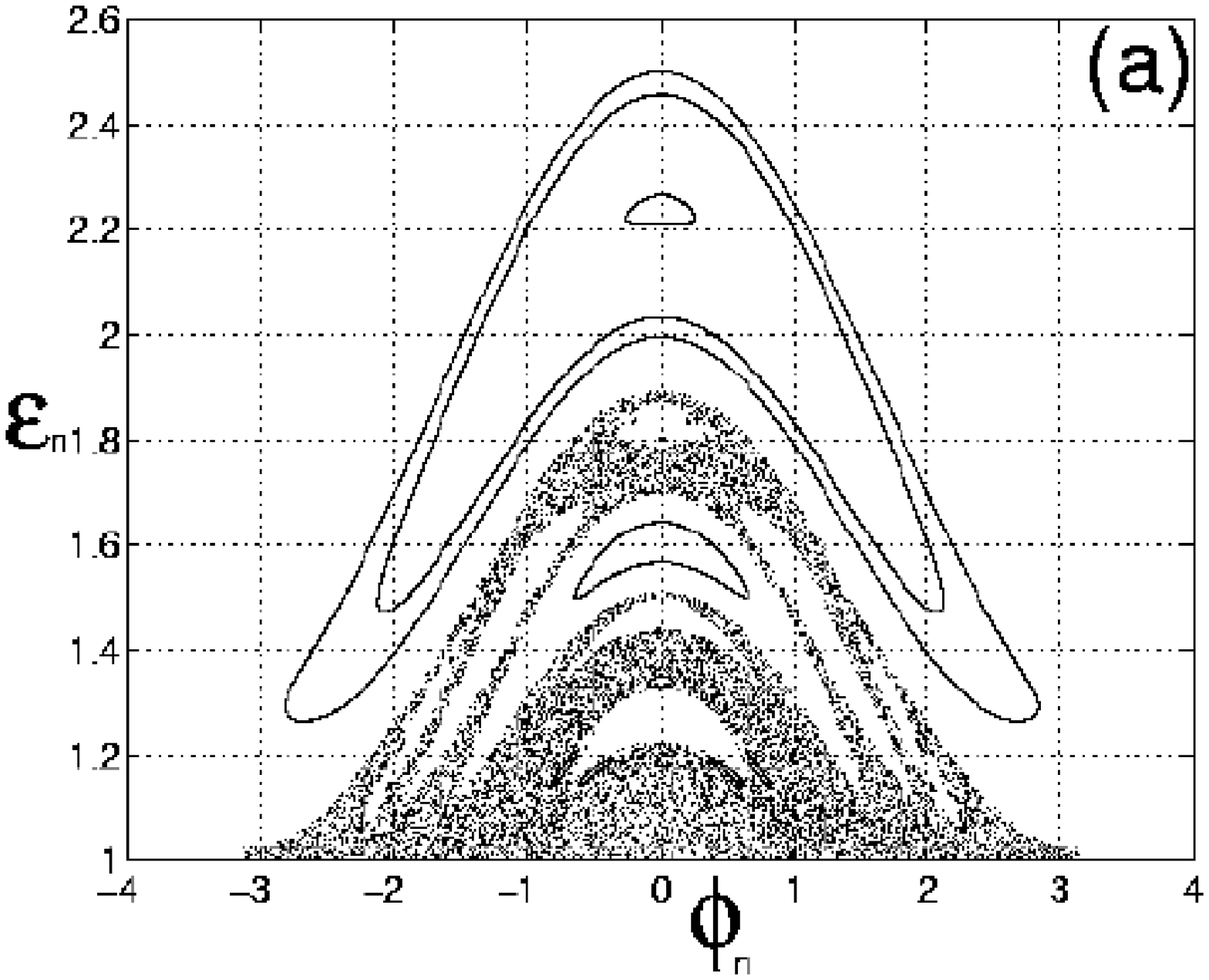}
\includegraphics[scale=0.45]{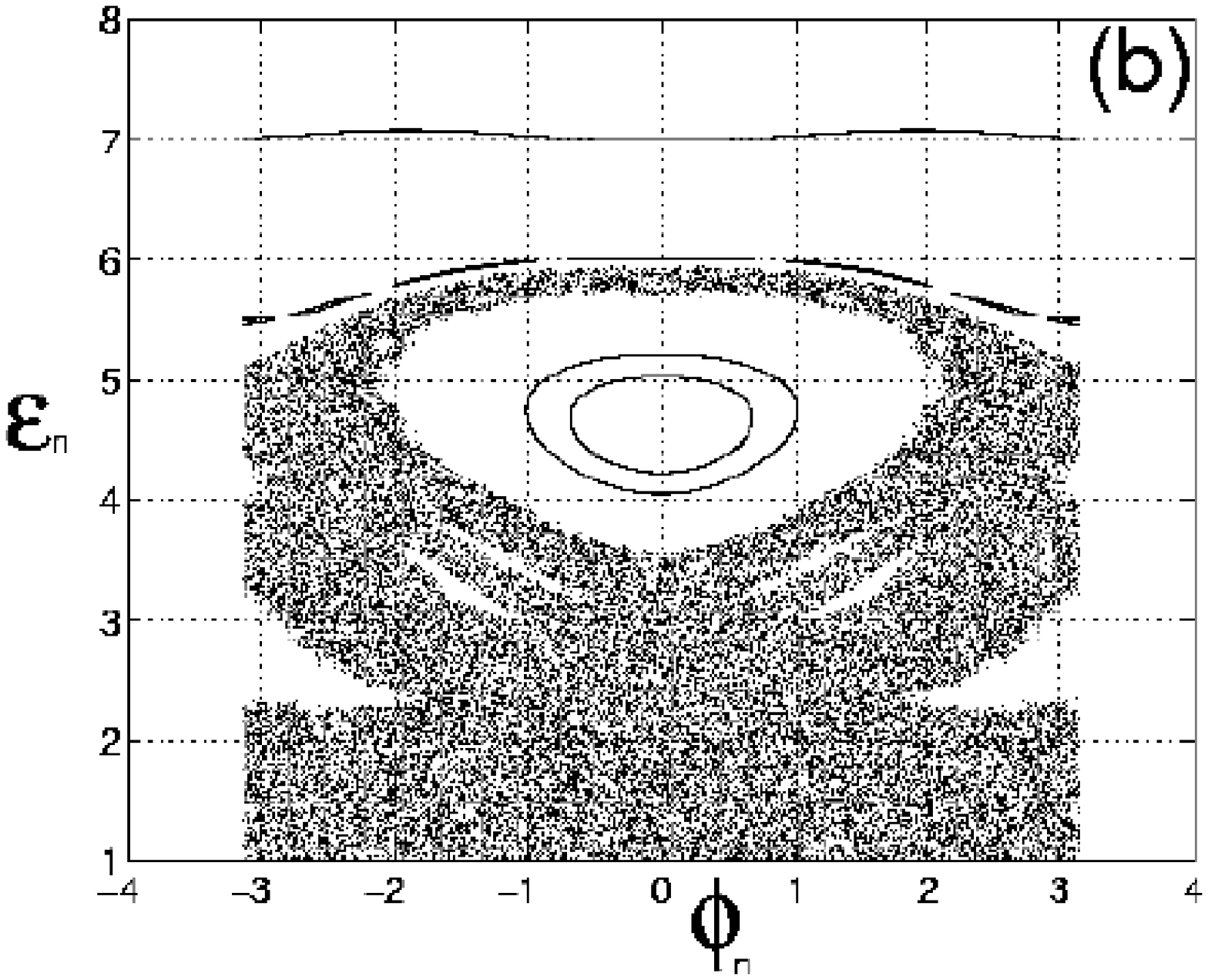}
\caption{Poincare plots for the oscillating chain of wells. a) $\eta_3=0.1$ and b) $\eta_3=10$. Same $\eta_1$ and
$\eta_2$ as in Fig. 15. }
\label{fig:16}
\end{figure}

\begin{figure}[!htb]
\centering
\includegraphics[scale=0.45]{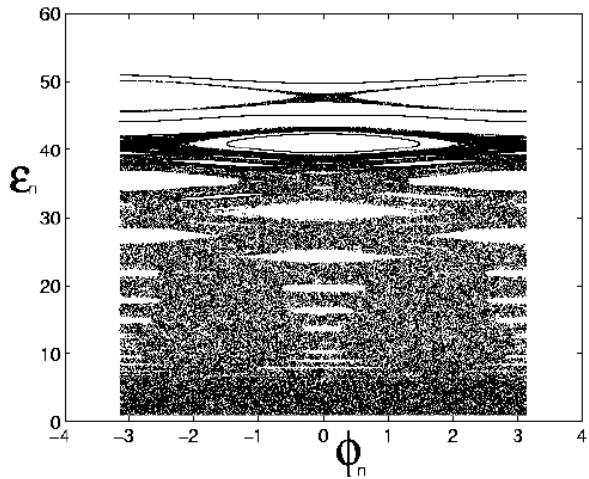}
\caption{Poincare plot for the oscillating well with $\eta_3=500$. Same $\eta_1$ and
$\eta_2$ as in Fig. 16. }
\label{fig:17}
\end{figure}

Another formula that will be useful below is obtained by substituting 
$\phi'_{m_n}=\phi_{m_n}-\frac{1}{2}\delta \phi_{m_n}$ and $ \Phi_{m_n}= \phi_{m_n} +\frac{1}{2}\delta \phi_{m_n}$
in (20) and expanding the cosine terms:

\begin{eqnarray}
\epsilon_{m_n} & = & \epsilon_0 +\eta_1[cos(\phi_{m_n} - \frac{1}{2}\delta \phi_{m_n})-cos \Phi_0)] + {} \nonumber \\
& & {} + 2 \eta_1 \sum_{i=1}^{n-1}sin \phi_{m_i} \ sin\frac{1}{2}\delta \phi_{m_i}
\end{eqnarray}

\begin{eqnarray}
\phi_{m_n}-\frac{1}{2}\delta \phi_{m_n}&=& \phi_{m_{n-1}} + m_n \Delta \Phi_{m_{n-1}} + {} \nonumber \\
 & & {} + \frac{1}{2}\delta \phi_{m_{n-1}}
\end{eqnarray}
 
Some features of the Poincare plots for small $\eta_3$  can be understood qualitatively by first
analyzing the unperturbed case ( $\eta_3=0$). If $\eta_3=0$ then $\delta \phi_{m_i}= 0$, 
$\Phi_{m_i}= \phi_{m_i}= \phi'_{m_i}$ for all $i$, and (21) and (22)  reduce to

\begin{eqnarray}
  \epsilon_{m_n} & = & \epsilon_0 + \eta_1( cos\phi_{m_n} -cos\phi_0), \\
  \phi_{m_n}  & = & \phi_{m_{n-1}} + m_n  \Delta \phi_{m_{n-1}} \nonumber \\
 & = & \phi_0 +\Delta \phi_0 \sum_{i=1}^{n}m_i .
\end{eqnarray} 
 
To obtain the last equation we used the fact that $\Delta \phi_{m_i}=\Delta \phi_0$ for all $i$ since
$\epsilon_{m_i}-\eta_1 cos\phi_{m_i}$ is the (adimensional) internal kinetic energy, which is conserved
for $\eta_3=0$ ( see discussion below Eq. (1)).  Eqs. (23) and (24) show that for $\eta_3 =0$ all orbits are
described by invariant 1D curves, the cosine curves $\epsilon_{m_n}(\epsilon_0, \phi_0)$. Recall that
for a given interaction $n$ with a well $\epsilon_{m_n} \geq 1$ but $\epsilon_{m'_n}< 1$ for $m'_n< m_n$.
It is clear from (23) that if $\epsilon_0 > 1+ 2 \eta_1$ then $\epsilon_{m_n} > 1$ and $ m_n=1$ for all $n$ and
hence the motion is of the {\it{translational}} type; the particle nevers changes direction. 
If $\epsilon_0=1+2\eta_1$ then the right hand side of (23) can be as small as one for some $n$, 
provided  $\phi_0=0$ and $\phi_{m_n}=k\pi, k,$ odd. 
This implies that $ n  \Delta\phi_0=n \eta_2/\sqrt{1+\eta_1}$ must be exactly
equal to $k\pi$, but for any $\eta_1$ in the range $(0,1)$ and any $\eta_2\geq 0$, there are some
integers $n$ and $k$ that approximate arbitrarily close this equality. Since $\epsilon=1$ is the 
energy threshold, above which the particle does not reflect, then it follows that the curve
$\epsilon= 1 +\eta_1 + \eta_1cos\phi$ defines the {\it{separatrix}} curve, separating translational orbits from
 {\it{librational}} orbits ( curves below the separatrix).  For initial conditions placed below the separatrix,
$m_n$ is in general larger than one. Since all curves, translational or librational, are invariant curves, there is
no possibility for a translational orbit to become a librational one, and viceversa. The three top orbits of Fig. 14a
are translational and the rest are librational.

Now consider the $\eta_3 >0$ case.   Notice that the last term of (21) produces, as a function of 
$\phi_{m_n}-\frac{1}{2}\delta \phi_{m_n}$, fluctuations from the invariant cosine curve. These fluctuations
can be small or large, depending on the value of $\eta_3$ and whether the terms in the sum add constructively
or destructively.  Consider the case $\eta_3<<1$. If in addition $\delta \phi_{m_i}<<1$ for all $i$ then the summation
 term can be approximated
by $\eta_1 \sum \ sin\phi_{m_i}\delta\phi_{m_i}$. The summed terms would yield
an appreciable deviation from the cosine curve if $sin\phi_{m_i}$ were to oscillate slowly even if each term in the sum
is small. This could happen for low energies
( c.f., (22) and (18)) and the orbit in the Poincare surface will appear as scattered points in the neighborhood
of the cosine curve. Similarly,  if $\epsilon_{m_i}$ is close to one,  the last term of (19) may
become the dominant term.
Thus even for a small $\eta_3$, the perturbation can be strong for initial conditions in the neighborhood of the
unperturbed separatrix or for initial conditions on low-lying unperturbed librational orbits, see Fig. 14b. 
This is the concrete mechanism in this system for the appearance of chaotic orbits
as $\eta_3$ becomes slightly larger than zero ( Figs. 14b and 15).

Further analysis on the development of chaos as $\eta_3$ is increased will be reported elswhere. Here
we shall only remark on some features observed in  our numerical results ( Figs-15-17). 
It is interesting to note, for example, that the "boomerang" like
stable island immersed in the chaotic sea of Fig. 15b ( at $\epsilon_n\approx 1.2$) 
corresponds to the period one stable fixed
point of the double well. In fact, the Poincare plot for orbits trapped between two
wells (see Fig.13) fits exactly within the abovementioned stable island
of the chain, and the heteroclinic tangle of the double well lies on the chaotic sea of the
chain. The elliptic islands above the lowest one most likely  correspond to period one stable fixed points
produced by trajectories trapped between three, four, and more wells.\\

Figures 15b to 17 show that as $\eta_3$ continues to increase, the chaotic area
increases, allowing the energy to diffuse to higher and higher values.
Thus if the energy is initially very small,  after some time it may become very large as long as diffusion in energy is not limited by KAM
curves.\\

In addition to energy diffusion, there is also diffusion in configuration space since the particle is interacting
chaotically with the  wells.  Figure 18a plots the well number vs. number of interaction for the same 
parameters as in Fig. 17. Notice the random-walk-like behavior. That is, the particle, starting at well number one, seems to travel randomly
forward and backward, returning repeatedly to the starting point. While this is happening, the energy 
of the particle oscillates wildly between one and approximately seven (see Fig. 18b). 
Then after approximately 55,000  interactions ( see Fig. 18b), the particle
travels only  in the negative direction during several interactions. This corresponds to a jump
in its energy and the particle diffuses  to higher energy  values ( see Fig. 18b). This behavior can be
partially  understood by examing the Poincare plot of Fig.17. Notice
that for energies around seven units, there are  some visible white areas indicating the existence of
some large,  very thin, islands of stability, where the orbits get stuck temporarily. For larger number of 
interactions the particle eventually returns to the original well, performing a random-like-walk and the energy
fluctuates over the whole range delimited by the KAM curve found around $\epsilon=42$.

\begin{figure}[!htb]
\centering
\includegraphics[scale=0.45]{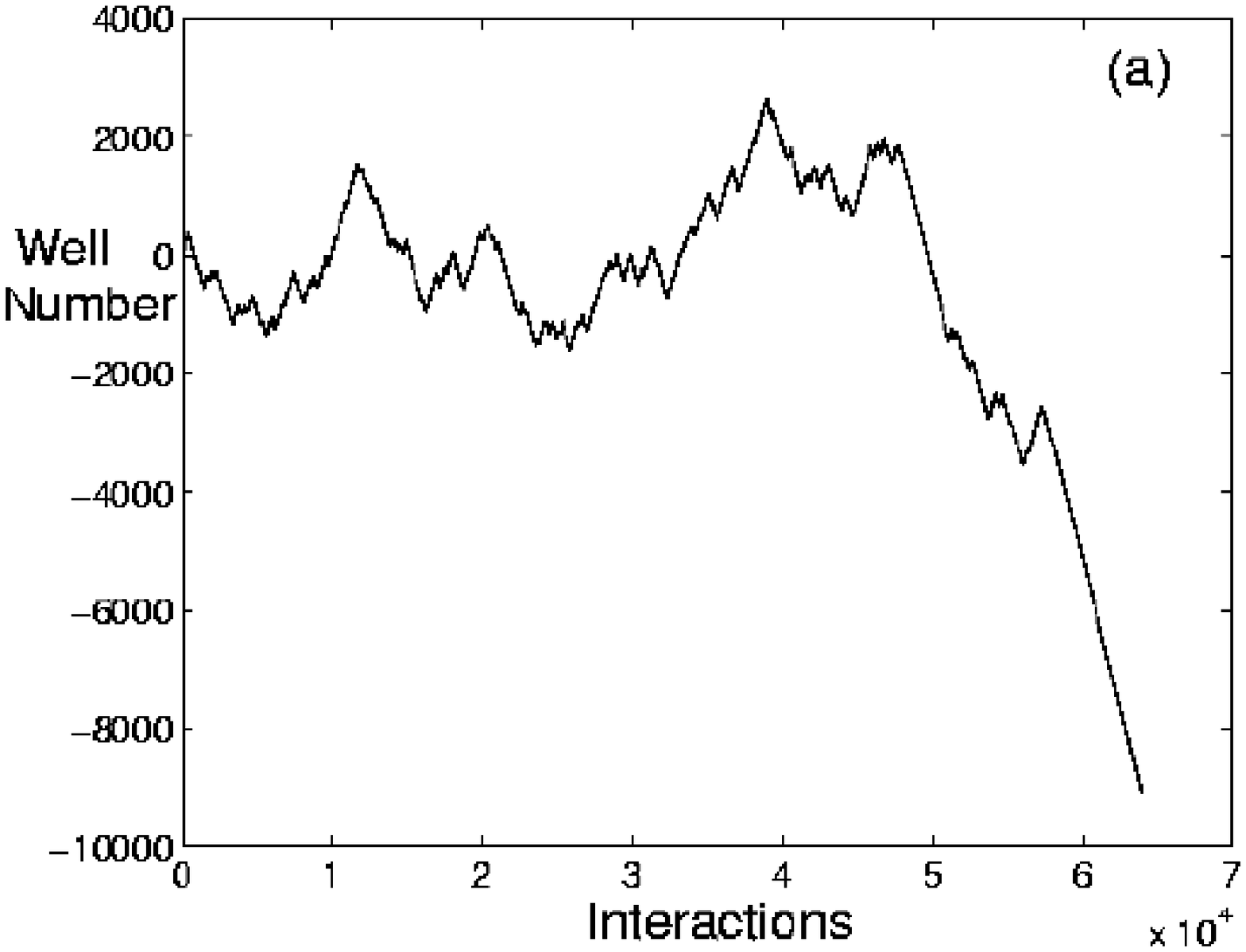}
\includegraphics[scale=0.45]{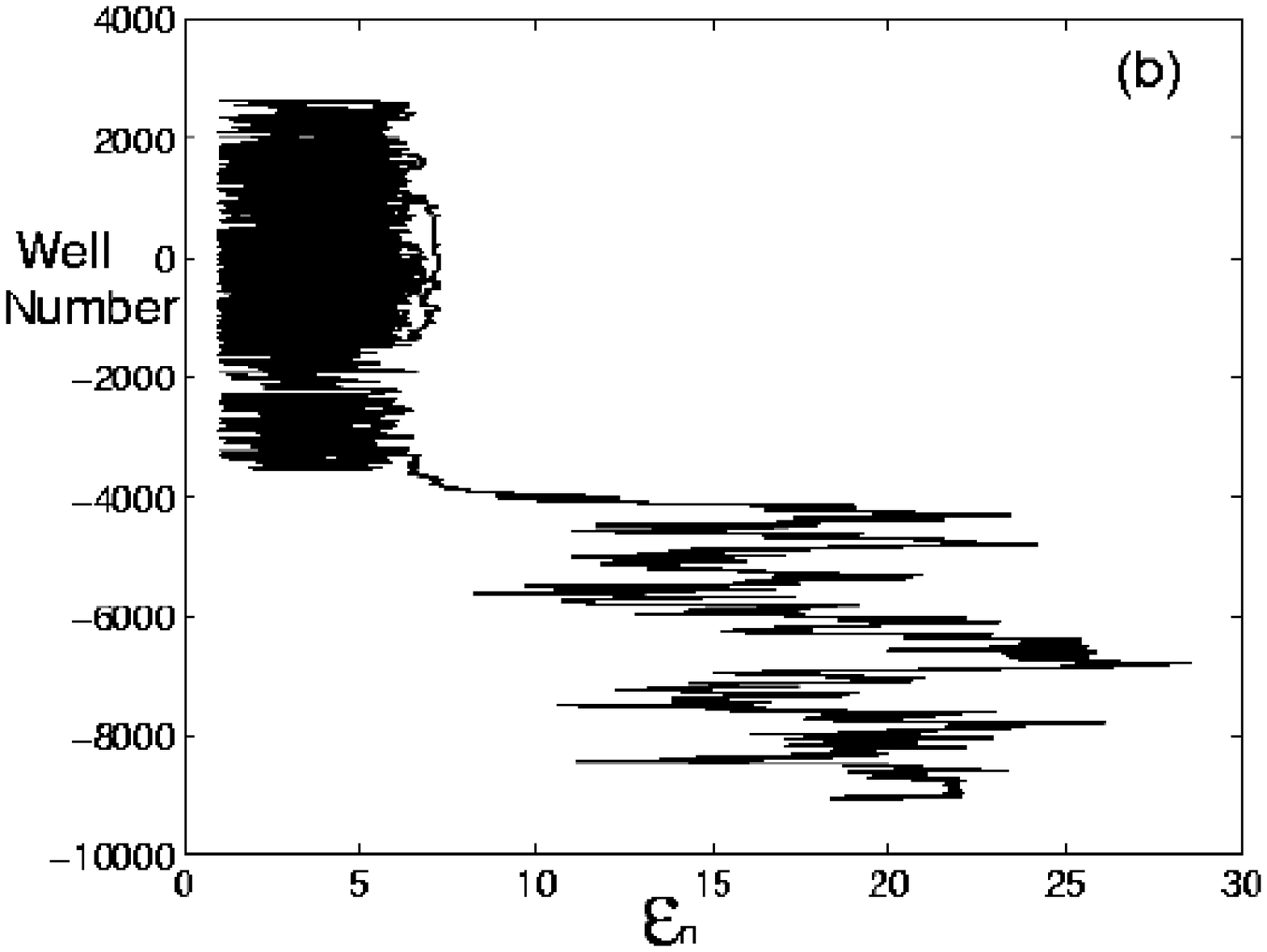}
\caption{Space and energy diffusion in the oscillating chain. a)  is diffusion in space, represented by the well
number and b) is diffusion in energy of the particle as it interacts with the various wells.  The parameters are
$\eta_1=0.5$, $\eta_3=16.45$, and $\eta_3=500$.}
\label{fig:18}
\end{figure}

\section{Conclusions}

In this paper we have described the motion of a classical  particle interacting
with one, two, and finally an infinite chain of oscillating square wells.  For the single well
we found that
although the scattering functions, such as final energy and number of collisions as a function
of initial conditions, may  yield complex behavior, extremely sensitive
to initial conditions ( hence unpredictable), the nonexistence of hyperbolic fixed points renders the scattering
non-chaotic. On the other hand, when the potential region is formed by two oscillating wells, the
scattering functions are truly fractal as expected since it was shown that there exists
topological chaos.  The dynamics of a particle interacting with a chain of oscillating wells
was shown to yield the transition to chaos as the separation between the wells
increases. Although there is chaotic diffusion in energy this is limited by KAM curves
and the diffusion in configuration space is random-like but with some peculiarities.
The scattering off one and two  oscillating square wells has been considered previously
by some authors  to model some microscopic and mesoscopic
systems but their treatments have been completely quantum mechanical. Here we treat the purely
classical description as a starting point to understand their classical-quantum correspondence.
 The oscillating chain represents a classical oscillating
Kronig-Penney model, particularly relevant to solid state systems, and it may help in the 
understanding of some quantum properties. For example, some localization of the wave function
 of an electron
in a one-dimensional crystal or in a superlattice in the presence of  some monochromatic 
field \cite {Bass} is expected to occur  since we have shown that the motion can be chaotic. This should 
manifest itself in the reduction of conductivity. Thus, there is
some similarity with the Anderson Model \cite {Anderson}.
The dynamics of a particle in  the oscillating chain is also similar to the Fermi 
accelerator model, with the additional nice feature that in the chain  there is diffusion in space as well
as in energy. Finally, as far as the energy is concerned  the oscillating chain is equivalent to a
single oscillating well inside a larger static infinite square well.  An oscillating square barrier inside a
larger static infinite square well has been recently considered in \cite {Jose}. 
\smallskip

\noindent{\bf Acknowledgments:}
We thank Luis Benet for illuminating discussions.
This work was partially supported by  CONACYT (Mexico) grant, No. 26163-E and
NSF-CONACYT, grant No. E120.3341


\end{document}